\documentclass[iop,twocolumn]{aastex631} 
\usepackage{amsmath,apjfonts}

\DeclareMathOperator{\sech}{sech}

\usepackage{textcomp}
\usepackage{gensymb}
\usepackage{hyperref}

\newcommand\xone{x_1}
\newcommand\xtwo{x_2}
\newcommand\xthree{x_3}
\newcommand\cs{c_s}

\newcommand\Msun{\; {M}_{\odot}}
\newcommand\Lsun{\; {\rm L}_{\odot}}
\newcommand\kms{\; {\rm km}\;{\rm s}^{-1}}

\newcommand\pc{\;{\rm pc}}

\newcommand\kpc{\;{\rm kpc}}

\newcommand\freq{\kms\kpc^{-1}}

\newcommand\cm{\;{\rm cm}}

\newcommand\Myr{\;{\rm Myr}}

\newcommand\MBH{M_{\rm BH}}

\newcommand\RCR{R_{\rm CR}}

\newcommand\C{{\mathcal C}}

\newcommand\simgt{\lower.5ex\hbox{$\; \buildrel > \over \sim \;$}}
\newcommand\simlt{\lower.5ex\hbox{$\; \buildrel < \over \sim \;$}}
\newcommand{\RNum}[1]{\uppercase\expandafter{\romannumeral #1\relax}}

\newcommand\oiii{\rm [O\:{\RNum{3}}]}
\newcommand\HI{\rm H\:{\RNum{1}}}

\graphicspath{{./}{figures/}}

\begin{document}

\title{Bar-driven Gas Dynamics of M31}


\author{Zixuan Feng}
\affiliation{Shanghai Astronomical Observatory, Chinese Academy of Sciences, 80 Nandan Road, Shanghai 200030, P.R. China}
\affiliation{School of Astronomy and Space Sciences, University of Chinese Academy of Sciences, 19A Yuquan Road, Beijing 100049, P. R. China}

\author{Zhi Li}
\affiliation{Shanghai Key Lab for Astrophysics, Shanghai Normal University, 100 Guilin Road, Shanghai 200234, China}
\affiliation{Department of Astronomy, School of Physics and Astronomy, Shanghai Jiao Tong University, 800 Dongchuan Road, Shanghai 200240, P. R. China}

\author{Juntai Shen}
\affiliation{Department of Astronomy, School of Physics and Astronomy, Shanghai Jiao Tong University, 800 Dongchuan Road, Shanghai 200240, P. R. China}  

\email{jtshen@sjtu.edu.cn}
\affiliation{Key Laboratory for Particle Astrophysics and Cosmology (MOE) / Shanghai Key Laboratory for Particle Physics and Cosmology, Shanghai 200240, P. R. China}

\affiliation{Shanghai Astronomical Observatory, Chinese Academy of Sciences, 80 Nandan Road, Shanghai 200030, P.R. China}

\author{Ortwin Gerhard}
\affiliation{Max-Planck-Institut f{\"u}r Extraterrestrische Physik, Giessenbachstrasse, D-85748 Garching, Germany}

\author{R. P. Saglia}
\affiliation{Max-Planck-Institut f{\"u}r Extraterrestrische Physik, Giessenbachstrasse, D-85748 Garching, Germany}
\affiliation{Universit{\"a}ts-Sternwarte München, Scheinerstr. 1, 81679 Munich, Germany}

\author{Matias Bla\~{n}a}
\affiliation{Institute of Astrophysics, Pontificia Universidad Católica de Chile, Av. Vicuña Mackenna 4860, 7820436 Macul, Santiago, Chile}
\affiliation{Max-Planck-Institut f{\"u}r Extraterrestrische Physik, Giessenbachstrasse, D-85748 Garching, Germany}

\author{Hui Li}
\affiliation{Department of Astronomy, Tsinghua University, Beijing 100084, P.R. China}
\affiliation{Department of Astronomy, Columbia University, New York, NY 10027, USA}

\author{Yingjie Jing}
\affiliation{National Astronomical Observatories, Chinese Academy of Sciences, 100101 Beijing, China}

\begin{abstract}
The large-scale gaseous shocks in the bulge of M31 can be naturally explained by a rotating stellar bar. We use gas dynamical models to provide an independent measurement of the bar pattern speed in M31. The gravitational potentials of our simulations are from a set of made-to-measure models constrained by stellar photometry and kinematics. If the inclination of the gas disk is fixed at $i = 77^{\circ}$, we find that a low pattern speed of $16-20\freq$ is needed to match the observed position and amplitude of the shock features, as shock positions are too close to the bar major axis in high $\Omega_{b}$ models. The pattern speed can increase to $20-30\freq$ if the inner gas disk has a slightly smaller inclination angle compared with the outer one. Including sub-grid physics such as star formation and stellar feedback has minor effects on the shock amplitude, and does not change the shock position significantly. If the inner gas disk is allowed to follow a varying inclination similar to the $\HI$ and ionized gas observations, the gas models with a pattern speed of $38 \freq$, which is consistent with stellar-dynamical models, can match both the shock features and the central gas features. 

\end{abstract}



\section{Introduction}
\label{sec:intro}

The neutral and ionized gas in the inner $10\arcmin\times10\arcmin$ ($2.3\kpc\times2.3\kpc$) of M31 exhibits strong non-circular motions that are likely caused by a rotating stellar bar. For example, \citet{sta_bin_94} proposed that gas flows within a barred potential can create the "face-on spiral" patterns seen in dust and ionized gas \citep{cia_etal_88} as well as observed gas velocity patterns in the inner region of M31 \citep{rub_for_71, bri_bur_84, bri_sha_84}. However, the comparison between their models and observations was relatively qualitative. Using a barred potential constrained with the $B$-band surface brightness profile, \citet{berman_01} constructed a series of hydrodynamical models for M31. Their best-fitting model successfully reproduced the observed line-of-sight velocities $V_{los}$ of CO along the disk major axis \citep{loi_etal_95}. \citet{ber_loi_02} further developed 3D models to explain off-axis CO observations in the southern half of M31 \citep{loi_etal_99}. These pioneer studies imply that the dynamical properties of the bar in M31, including its pattern speed and quadrupole moment, may be independently determined through gas dynamical models. 

The VIRUS-W IFU study, presented by \citet{opi_etal_18}, covers the central bulge region and part of the stellar disk in M31. These observations have improved our understanding of the inner structures of M31 by revealing comprehensive stellar and gas kinematics. Many kinematic features, such as twisted zero-velocity lines in both the stellar and gas velocity fields, the correlation between the Gauss-Hermite moment h3 and $V_{los}$ in the bulge region, and highly irregular gas features displaying strong non-circular motions, suggest the presence of a stellar bar in M31. \citet{bla_etal_18} have constructed made-to-measure (m2m) models to fit the infrared-band photometry in \citet{bar_etal_06} and the stellar kinematics obtained by \citet{opi_etal_18}. These m2m models are based on the N-body models created by \citet{bla_etal_17}, which consist of a classical bulge and a boxy/peanut bulge with masses of approximately 1/3 and 2/3 of the total bulge mass, respectively. The bar in their best-fitting models has a half-length of $a_{bar} = 4 \kpc$ and a pattern speed of $\Omega_{b} = 40\pm 5\freq$. While the m2m models have provided more constraints on the gravitational potential of M31, hydrodynamical simulations are still necessary to determine whether the gravitational potential can reproduce the observed gas features.

The global gas morphology of M31 is not only influenced by the gravitational potential, but also by its interaction history. \citet{blo_etal_06} proposed that a head-on collision between M32 and M31, which occurred about 210 million years ago with a mass ratio of 1:10, could explain the two off-centered rings observed in the IRAC $8\;\rm \micro m$ image of M31. The collision in their model produced an outer gas ring moving at radial velocities of about 10 km/s and an inner gas ring with a more "face-on" morphology. More recently, \citet{ham_etal_18} used hydrodynamical simulations of a major merger to explain the steep age-velocity dispersion relation found in the stellar disk \citep{bha_etal_19} and the enhancement of star formation about 2-4 billion years ago in M31 \citep{wil_etal_15}. They estimated that a 1:4 merger occurred about 1.8-3 billion years ago, producing the Giant Stream in the halo of M31 with a pericenter radius of approximately 32 $\kpc$. In their model a 10-kpc ring and a central bar formed after the coalescence of the nuclei. The 10-kpc ring in their model is consistent with observations by \citet{lew_etal_15}, and the central bar roughly reproduces the observed photometry of the bulge region. These simulations all used a bar to explain the isophotal twist in the bulge region of M31 \citep{lindbl_56}, but the bar parameters in these models were not constrained by the central stellar and gas kinematics. 

Shock features on the leading side of the bar are typical signatures of barred galaxies. Gas loses a significant amount of angular momentum as it crosses these shocks, producing sharp velocity jump features on position-velocity diagrams (PVDs). These shock features are highly sensitive to the mass distributions and bar parameters \citep{athana_92, kim_etal_12b, li_etal_15}, which can break the disk-halo degeneracy and provide constraints on the bar pattern speeds, as in previous research for NGC~4123, NGC~1365, and NGC~1297 \citep{wei_etal_01a, wei_etal_01b, zan_etal_08, fra_etal_17}. By identifying these velocity jump features (shock features) on PVDs of $\oiii$ and $\HI$, \citet{fen_etal_22} found that they are distributed regularly on a large scale ($\kpc$) and mainly located on the leading side of the bar in M31. The shock features follow a typical pattern of bar-driven gas flow, with the largest velocity jumps exceeding $170\kms$ on the gas PVDs.

\begin{figure*}[t!]
\centering
\includegraphics[width=0.8\textwidth]{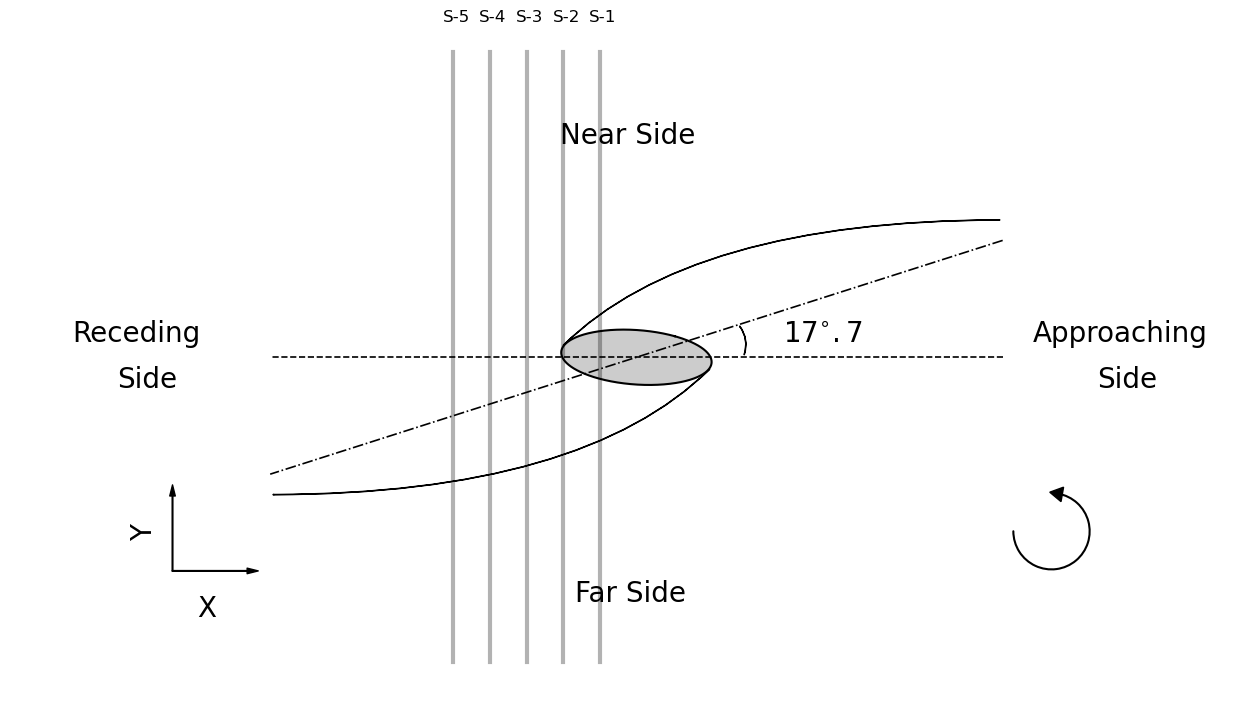}
\caption{A schematic diagram showing homogeneously positioned pseudo-slits (vertical solid lines) overlaid on the expected bar-driven gas inflow in M31. Slit numbers are labeled on the top of the pseudo-slits for later reference. Shock features on PVDs of these five pseudo-slits are used to give constrains on gas models. The horizontal dashed line represents the line-of-nodes of the stellar disk in M31. The dash-dotted line indicates the bar major axis in \citet{bla_etal_18}, which is deviated from the dashed line by $\sim 17.7^{\circ}$. Black curves represent the expected bar-drive shock features on the leading side of the counterclockwise rotating bar. Please note that the inner ellipse represents a projected gas nuclear disk, instead of the bar of M31. The figure is adapted from \citet{fen_etal_22}. We show it here to help readers know the meaning of the slit number and make this article more self-sufficient. }
\label{fig:scheme_slit_position}
\end{figure*}

Our aim is to use the well-constrained gravitational potential presented in \citet{bla_etal_18} to run hydrodynamical simulations systematically and constrain the bar pattern speed with the identified $\oiii$ shock features. \citet{fen_etal_22} constructed a series of pseudo-slits in the central $20\arcmin\times10\arcmin$ ($4.6\kpc \times2.3\kpc$) region of M31 with a slit width of $1.2\arcmin (\sim274\pc)$. The pseudo-slits were positioned to be roughly perpendicular to the bar major axis of M31, which would reveal shock features most clearly on gas PVDs. \citet{fen_etal_22} found that evident shock features mostly locate on the far side of M31, although weaker shock features can be identified on both sides. Several $\oiii$ shock features were found near the end of the bar, however, these features were close to the edge of the observational coverage of $\oiii$ data and the relatively low data quality could not resolve the velocity jump clearly. For a comparison, we also found the $\HI$ shock features were not suitable to constrain gas models due to the insufficient resolution in $\HI$ data. We then constrained our gas models using the five clearest shock features in the $\oiii$ data on the far side of M31, which were identified in five pseudo-slits S-1 to S-5 located at $X = -1.2\arcmin$ to $-6.0\arcmin$ on the disk major axis of M31. The homogeneously constructed pseudo-slits are sufficiently representative of the shock region as a whole, as shown in Figure~\ref{fig:scheme_slit_position}. We restrict the comparison in the shock region rather than all the detailed gas features for several reasons: 1) Gas features in our models are bi-symmetric, while there are some asymmetries in the velocity field of $\oiii$. 2) The inner region of M31 shows a spiral pattern with an inclination possibly smaller than the stellar disk, which a 2D model cannot reproduce. 3) If a recent head-on collision indeed occurred, as suggested by \citet{blo_etal_06}, it would strongly perturb the inner gas disk. Further improvement in a tidal interaction scenario is discussed in \S~\ref{sec:tidal_interaction_scenario}.

The paper is structured as follows: In \S~\ref{sec:model_setup}, we introduce the model settings. In \S~\ref{sec:compare_model_obs}, we describe the method used to determine the goodness of fit. In \S~\ref{sec:search_for_bestfit_model}, we searched for best-fitting models with varying bar pattern speeds, gas sound speeds, inclinations, and gravitational potentials. We also show the shock features in several representative models. \S~\ref{sec:smuggle_model} presents the results obtained with a more sophisticated gas model. In \S~\ref{sec:model_varying_inclination} we investigate whether a gas model with a tilted inner disk helps reduce the pattern speed discrepancy between gas and stellar dynamical models. We mainly discuss other central gas features and the possible systematic uncertainties in gas and stellar dynamical models in \S~\ref{sec:discussion}. Finally, we provide a brief summary of our results in \S~\ref{sec:conclusion}.

\section{Model setup}
\label{sec:model_setup}

\subsection{Numerical scheme}
\label{sec:initial_setting}
We present hydrodynamic simulations with the latest version of the MHD code Athena++ \citep{sto_etal_19, sto_etal_20}. We solve the Euler equations on a uniform Cartesian grid with $2048 \times 2048$ cells, covering a simulation domain with a box size of $30\kpc$ along the $x$ and $y$ directions. This setting corresponds to a spatial resolution of approximately $15\pc$. Our initial gas surface density setting follows an exponential profile:

\begin{equation}
\Sigma_{\rm gas}(R) = \Sigma_{0}\exp(-R/R_{d}),
\end{equation}
here $\Sigma_{0} = 76 \Msun {\pc}^{-2}$ and $R_d = 6.0 \kpc$, giving the total gas mass of $\sim 1.2 \times 10^{10}\Msun$. The baryonic mass within $R = 15\kpc$ in the m2m models of \citet{bla_etal_18} is $\sim 7.5 \times 10^{10}\Msun$. The gas mass corresponds to a gas fraction of $\dfrac{1.2 \times 10^{10}\Msun}{7.5 \times 10^{10}\Msun} \sim 16\%$ inside $R = 15\kpc$. The numbers of $\Sigma_{0}$ and $R_d$ are consistent with the $\HI$ observations by \citet{bra_etal_09}. Initially, the gas disk is set in circular motions, with centrifugal force balancing the azimuthally averaged gravitational force. To avoid transients, we slowly ramp up the non-axisymmetric component of the gravitational potential. We achieve this by linearly decreasing the fraction of the axisymmetrized potential from 1.0 to 0, and increasing the fraction of the non-axisymmetric potential from 0 to 1.0 over 100 $\Myr$, as in previous studies \citep{sor_etal_15a, li_etal_22}. 

\begin{figure*}[t!]
\includegraphics[width=\textwidth]{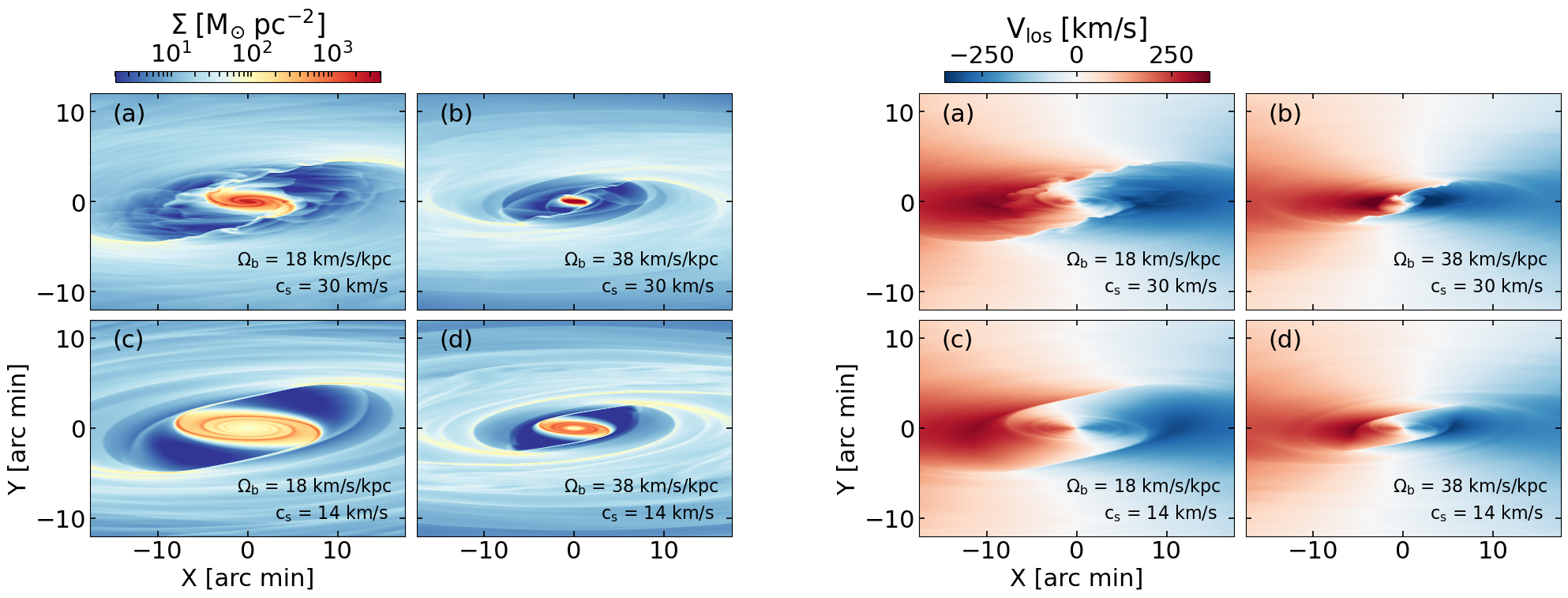}
\caption{\textit{Left panels}: Gas surface density of four characteristic models with JR804 potential. Snapshots are taken at $T = 800\Myr$. The gas surface densities are projected with an inclination of $77^{\circ}$ and a bar angle $\phi_b$ of $54.7^{\circ}$. The disk major axis (inclination axis) is along the $X$-axis. \textit{Right panels}: Similar to the left panels but for the distribution of line-of-sight velocity.}
\label{fig:4model_SD_comparison}
\end{figure*}

\begin{table*}
\caption{Main parameters of gas simulations}
\label{tab:par_m2m_models}
\begin{center}
\begin{tabular}{c c c c c c c c c}
\hline
\hline
{Gas model group} & {Model potential} & {$M_{\star}^{CB}$} & {$M_{\star}^{BPB}$} & {$M_{DM}^{B}$} & {$\Omega_{b}$ test range} & {Reference $\Omega_{b}$}  & {$\cs$ test range} & {Number of} \\
{} & {} & {($10^{10}\Msun$)} & {($10^{10}\Msun$)} & {($10^{10}\Msun$)} & {($\freq$)} & {($\freq$)} & {($\kms$)} & {models} \\
\hline
{GxJR804} & {JR804} & {1.18} & {1.91} & {1.2} & {14-50} & {40} & {10-38} & {127} \\
\hline
{GxKR241} & {KR241} & {1.16} & {1.82} & {1.0} & {14-50} & {40} & {10-38} & {127} \\
\hline
{GxJROb} & {JR924} & {1.22} & {1.87} & {1.2} & {50} & {55} & {10-38} & {8}\\
~ & {JR844} & {1.20} & {1.90} & {1.2} & {46} & {45} & {10-38} & {8}\\
~ & {JR804} & {1.18} & {1.91} & {1.2} & {38, 42} & {40} &  {10-38} & {16}\\
~ & {JR764} & {1.15} & {1.93} & {1.2} & {34} & {35} & {10-38} & {8}\\
~ & {JR724} & {1.10} & {1.94} & {1.2} & {26-32} & {30} &  {10-38} & {22}\\
~ & {JR644} & {1.02} & {2.04} & {1.2} & {14-24} & {20} &  {10-38} & {65}\\
\hline
\end{tabular}
\end{center}
\tablenotetext{}{Note that the m2m potentials are given by \citet{bla_etal_18}. $M_{\star}^{CB}$, $M_{\star}^{BPB}$, and $M_{DM}^{B}$ are calculated as the mass within 3.2 $\kpc$. Each model corresponds to one point in Figure~\ref{fig:overview_SMHD_I77}.}
\label{tab:criteria_shock_feature}
\end{table*}

We use an isothermal equation of state (EoS) with the \textit{effective} sound speed $c_{s}$ as a free parameter. We have verified that the main gas features and kinematics in the 2D isothermal models are similar to those in the more sophisticated models. We also tested a more sophisticated gas model of M31 in \S~\ref{sec:smuggle_model}. The sound speed $c_{s}$ reflects the level of turbulence in the interstellar medium \citep{tam_etal_09}. The assumption of isothermal EoS allows us to focus on the response of gas to the non-axisymmetric gravitational potential by ignoring the local physics like cooling/heating, self-gravity, star formation, and stellar feedback etc. For its simplicity and computational efficiency, isothermal EoS is well suited to systematically explore large parameter spaces and further give constraints on the properties of large-scale dynamical structures. Previous gas models using isothermal EoS have explained many observational gas features in the Milky Way, from gas kinematics of the Central Molecular Zone \citep[CMZ;][]{rid_etal_17, li_etal_20} to multiple gas structures driven by the bar and spiral arms on the longitude-velocity $(l-v)$ diagrams \citep{li_etal_16}. The more recent isothermal gas models in \citet{li_etal_22} not only explained the gas features on the $l-v$ diagram but also the non-circular motions of masers. These simulations gave further constraints on the overall mass distribution and bar pattern speed of the Milky Way. Additionally, isothermal models have been used to break the degeneracy of the baryonic matter and the dark matter, and determine the mass-to-light ratios in NGC~4123 \citep{wei_etal_01a, wei_etal_01b} and NGC~1365 \citep{zan_etal_08}.

\subsection{Setup of the gravitational potential}
\label{sec:grav_potential}

Our gravitational potential is determined by the made-to-measure (m2m) models constructed by \citet{bla_etal_18}. These models are well constrained by several observed quantities. The mass distribution is determined by the 3.6 $\rm \micro m$ photometry from IRAC observations \citep{bar_etal_06} and the $\HI$ rotation curve of the disk \citep{che_etal_09, cor_etal_10}. The VIRUS-W IFU survey \citep{opi_etal_18} provides data on the stellar kinematics in the bulge region, allowing for accurate constraints on the central dynamics.

We mainly use the two best-fitting m2m models JR804 and KR241 in \citet{bla_etal_18} as the basis of our gravitational potential. JR804 and KR241 use the Einasto and NFW dark matter halo profiles, respectively, and have the same fiducial bar pattern speed of $\Omega_{b} = 40\pm5 \freq$, placing the corotation radius at $R_{CR} = 6.5 \pm 1 \kpc$. Additionally, Hubble Space Telescope spectroscopy observations confirm the existence of a supermassive black hole in the nucleus of M31 \citep{ben_etal_05}. Therefore, we include a Plummer sphere to represent the gravitational potential of the central black hole: 

\begin{equation}
\Phi(r) = -G\MBH\dfrac{1}{\sqrt{r^2+a^2}},
\end{equation}

\noindent here $a = 10\pc$. We use black hole mass of $M_{\rm BH} = 2\times10^{8}\Msun$ as suggested by \citet{ben_etal_05}.

We simulate three groups of gas models that use different gravitational potentials: 

$-$ GxJR804: the gas model group that uses the JR804 potential.

$-$ GxKR241: the gas model group that uses the KR241 potential.

$-$ GxJROb: the gas model group that uses the series of m2m potentials with different $\Omega_b$. We consider five additional gravitational potentials from other m2m models as well (see table~\ref{tab:criteria_shock_feature}). These m2m models are fitted to observational data with the Einasto dark matter halo profile and a 3.6 $\rm \micro m$ mass-to-light ratio of $\Upsilon_{3.6} = 0.72\pm0.02\Msun \Lsun^{-1}$, but use different bar pattern speeds of $\Omega_{b} =$ 20, 30, 35, 40, 45, and 50 $\freq$, respectively. The mass ratios of classical bulges to boxy/peanut bulges in these models decrease as the pattern speed decreases. The rotation curves of these models differ slightly due to variations in their mass distributions. 

Overall, we simulate around 380 models to explore the gas evolution in different potentials with different bar pattern speeds and sound speeds. The gas models cover a large 2D parameter space with pattern speeds $\Omega_{b}$ in the range of $14 - 50\freq$ and sound speeds $c_{s}$ in the range of $10 - 38\kms$. The main parameters of gas simulations are listed in table~\ref{tab:criteria_shock_feature}.

\section{Comparison of models with observations}
\label{sec:compare_model_obs}

\subsection{Overall gas morphology and kinematics}

We adopt the distance to M31 to be $785\pm25\kpc$ \citep{mcc_etal_05} (at this distance $1\arcmin = 228\pc$), the inclination of the stellar disk of M31 to be $i = 77^{\circ}$ \citep{wal_sch_87}. The bar in the m2m models deviates from the disk major axis by a bar angle of $\phi_{b} = 54.7^{\circ}$ in the face-on view. After projection $\phi_{b}$ results in a deviation of $\phi_{b,\;proj} = \arctan(\tan\;\phi_{b}\times \cos\;i) = 17.7^{\circ}$ between the disk major axis and the bar major axis. We adopt the bar angle to be $\phi_{b} = 54.7^{\circ}$ in our gas models following \citet{bla_etal_18}.


The typical bar-driven gas flow patterns in the hydro simulations have been described in \citet{fen_etal_22}. Here we briefly introduce the gas substructures and present how they change with different $\Omega_{b}$ and $c_{s}$ in Figure~\ref{fig:4model_SD_comparison}. The models use the JR804 potential and are projected with a bar angle of $\phi_{b} = 54.7^{\circ}$ and an inclination of $i = 77^{\circ}$. The left panels of Figure~\ref{fig:4model_SD_comparison} present the gas surface density for the four characteristic models. The left and right columns present models with bar pattern speeds of $\Omega_{b} = 18\freq$ and $38\freq$, corresponding to the corotation radius of $R_{CR} \sim 14.1\kpc$ and $6.4 \kpc$, respectively. Models in the top and bottom rows use sound speeds of $c_{s} = 30\kms$ and $14 \kms$, respectively. The barred potential generates a nuclear ring with a high surface density near the center. A pair of shocks emerge from the outer rim of the nuclear ring, extending outwards on the leading side of the bar. A few bar-driven spiral features with a trailing shape appear near the end of the shocks. As the pattern speed increases, the shocks move closer to the bar major axis, resulting in a smaller nuclear ring. Increasing the sound speed produces more turbulent gas features, as shown in panels (a) and (b). A higher $c_{s}$ also results in smaller and denser nuclear rings, as well as shock features closer to the bar major axis. The overall gas patterns of our models are similar to the previous studies \citep{athana_92, kim_etal_12a, li_etal_15}. 

The right panels of Figure~\ref{fig:4model_SD_comparison} show the distribution of line-of-sight velocities for the four characteristic models. The red and blue colors indicate that the gas is moving away from us and toward us, respectively. Within the nuclear ring in all four models, the gas exhibits disk-like nearly circular motion. However, as we move farther out, the gas motion is dominated by the bar, resulting in strong non-circular motions (e.g. the high-velocity features at $X \sim \pm 10 \arcmin$ near the disk major axis in panel a). The loss of angular momentum as gas crosses the shocks is indicated by the sharp transition between red and blue colors. 

\subsection{Defining the goodness of fit}
\label{sec:define_goodness_of_fit}
In the VIRUS-W IFU study of M31 \citep{opi_etal_18}, the pixels are Voronoi binned to reach signal-to-noise ratios greater than 30. All pixels within a given bin share the same line-of-sight velocity $V_{los}$. The positions of the bins are described using a coordinate system introduced in \citet{fen_etal_22}, where the $X$-axis is aligned with the disk major axis and the $Y$-axis is aligned with the disk minor axis. We represent the position $(X_{bin}$, $Y_{bin})$ of each bin using the average $(X$, $Y)$ positions of the pixels within it. We begin by interpolating the projected velocity field of our gas models to obtain $V_{los}$ at all pixels. To generate a velocity map similar to the observations, we then compute the average $V_{los}$ of the pixels within each Voronoi bin to obtain the velocity of the bin $V_{bin}$. This process is repeated until all of the bins are accounted for. 

Following the method of constructing pseudo-slits in \citet{fen_etal_22}, we position five pseudo-slits perpendicular to the disk major axis at $X$ positions of $-1.2\arcmin$, $-2.4\arcmin$, $-3.6\arcmin$, $-4.8\arcmin$, and $-6.0\arcmin$, and label them as S-1, S-2, S-3, S-4, and S-5, respectively. The pseudo-slits have a width of $\Delta X = 1.2\arcmin$. We plot the gas PVDs along these pseudo-slits and obtain several boxcar smoothed curves representing the gas features. We find that models with $\Omega_{b}\leq 14\freq$ and $c_{s} < 18 \kms$ produce shock features outside the observational coverage, and thus can be excluded. To make a reasonable comparison for all the models, we focus on the similarity of velocity features within the region of $\oiii$ shock features. We define a comparison region as a window with a width of $\Delta Y = 2.4\arcmin$ around each $\oiii$ shock feature at $Y_{shock}$. This width is chosen to include the shock features in both models and observations. A slightly different width does not affect our main result. Note that we do not identify shock features in models directly, but compare their velocity features with $\oiii$ in the comparison region ($| Y-Y_{shock}| < 0.5\times\Delta Y$).

To assess the similarity between model features and $\oiii$ data, we use a modified version of the feature-comparing method described in \citet{sor_mag_15}. The procedure involves the following steps:

    1. Present the boxcar smoothed curves of gas PVDs on a grid with a spacing of $\Delta p = 0.05\arcmin$ and $\Delta v = 8\kms$. These values are empirically chosen to extract feature differences. Other values for $\Delta p$ and $\Delta v$ are tested and found to have no significant impact on the main result.
    
    2. Utilize the thinning algorithm developed by \citet{zha_sue_84} to make the curves thinner, reducing their width to one pixel.
    
    3. Use the Symmetrized Modified Hausdorff Distance \citep[SMHD,][]{sor_mag_15} to quantitatively measure the similarity of the curves between models and observations.

We represent the curves using binary images, where the pixels with a value of one indicate the presence of a feature. Suppose we have two binary images, A and B, representing the features in the model and the observation, respectively. We use $a_{i}$ and $b_{j}$ to represent the pixels with a value of one in A and B, respectively. For each $a_{i}$, we calculate its distance to each $b_{j}$ as follows:

\begin{equation}
d(a_{i},b_{j})= \dfrac{| p(a_{i})-p(b_{j})|}{\Delta p} + \dfrac{| v(a_{i})-v(b_{j}) |}{\Delta v},
\end{equation}

here $\Delta p$ and $\Delta v$ represent the spacing of the grids in A and B, respectively. We then find the $b_{j}$ closest to $a_{i}$ and record their distances as $d_{i} = min_{j}(d(a_{i},b_{j}))$. We repeat this step for all the pixels in A. The Modified Hausdorff distance is defined as the sum of $d_{i}$ for all $a_{i}$ in A:

\begin{equation}
{\rm MHD_{A}}  \equiv \sum_{i=1}^{N}d_{i}
\end{equation}

Then we use the same approach but acting on B to obtain $\rm MHD_{B}$. The SMHD is defined as: 

\begin{equation}
\label{equ:SMHD}
{\rm SMHD} \equiv \dfrac{\rm MHD_{A}}{2N}+\dfrac{\rm MHD_{B}}{2M},
\end{equation}

here $N$ and $M$ represent the total number of $a_{i}$ and $b_{j}$ in A and B, respectively. 

We repeat these steps for all five pseudo-slits from S-1 to S-5, resulting in a total of five SMHDs. We use the average of the five SMHDs to determine the goodness of fit. Although the models reach a quasi-steady state after around two bar rotation periods, they still exhibit transient features, such as gas clumps moving inward along shocks. These transient gas features can affect the positions of shocks and, consequently, the SMHD. To reduce the impact of transient shifts in shock positions and obtain a better quantification of the goodness of fit, we calculate the time average of SMHDs of 40 snapshots within $800-1000 \Myr$, which are uniformly spaced by $\Delta T = 5\Myr$.

\begin{figure*}[t!]
\includegraphics[width=\textwidth]{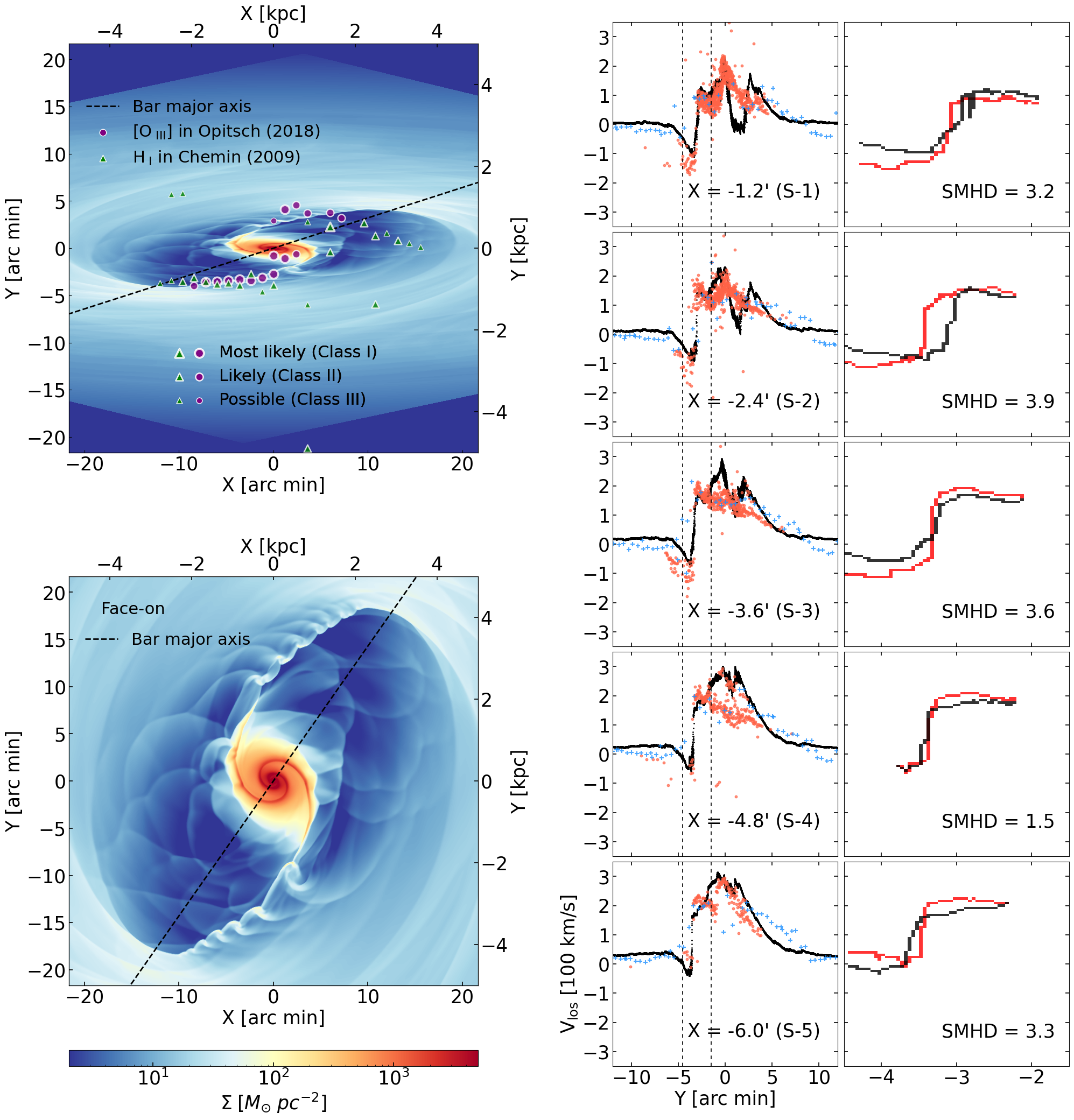}
\caption{\textit{Left panels}: Gas surface density of the best-fitting model projected to the sky (top) and in the face-on unprojected view (bottom) at $T = 950\Myr$. The model uses a bar pattern speed of $\Omega_{b} = 18\freq$ and a sound speed of $c_s = 34\kms$. The projection uses an inclination of 77$^{\circ}$. The disk major axis (inclination axis) is along the X-axis. Purple circles and green triangles indicate the shock positions in $\oiii$ and $\HI$, respectively. Large, medium, and small markers represent the Class I, Class II, and Class III shock features identified in \citet{fen_etal_22}, respectively. The dashed lines indicate the bar major axis in the model. \textit{Right panels}: Gas PVDs (first column) and identified shock features (second column) in the best-fitting model and their comparisons with the observations. The $x$-axis indicates the vertical distance from the disk major axis. The vertical axis indicates the line-of-sight velocities. Black lines represent the PVDs of the model along 5 pseudo-slits, with their positions shown in the bottom right corner. Data of $\oiii$ and $\HI$ are indicated by red and blue markers, respectively. The second column shows a zoom-in view of PVDs within the shock region, which is indicated by the vertical dashed lines in the first column. Black and red curves in the second column represent the identified shock features in the model and the $\oiii$ data, respectively. The similarity between the two features is quantified by the SMHD (Equation~\ref{equ:SMHD}) shown in the bottom right corner.}
\label{fig:KR241_bestfit_SD_pv}
\end{figure*}

\begin{figure*}[t!]
\includegraphics[width=\textwidth]{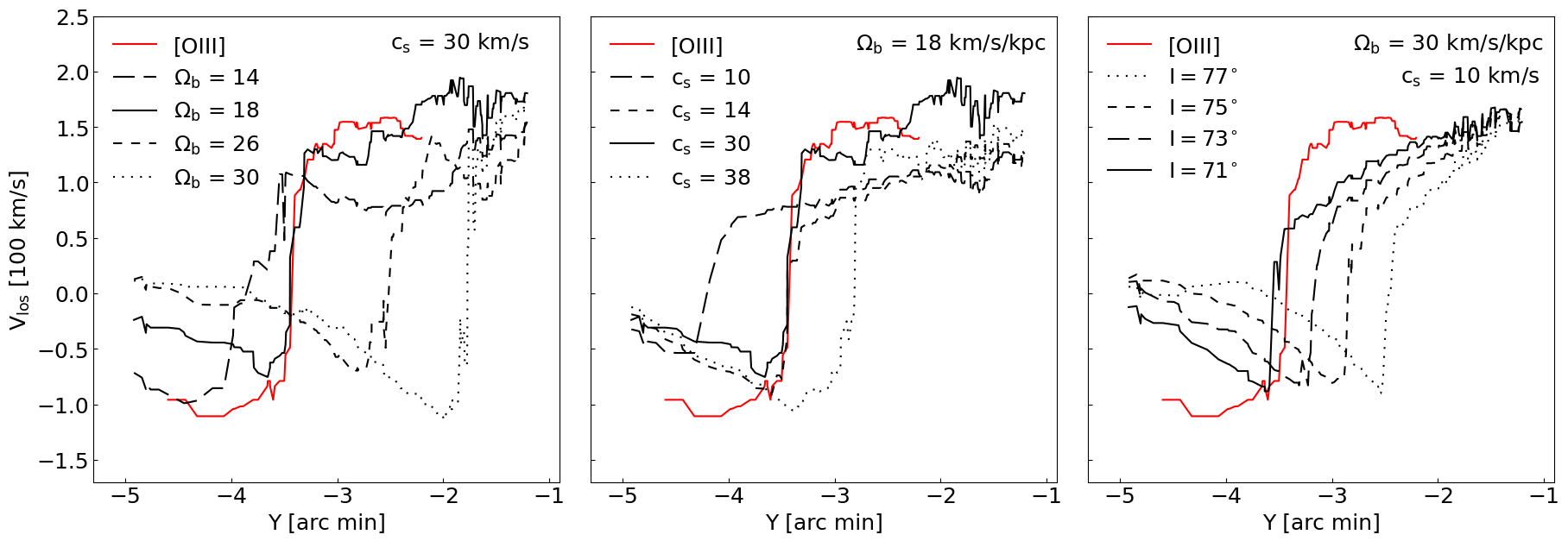}
\caption{\textit{Left panel:} The red curve represents the identified shock feature in $\oiii$ in a pseudo-slit at $X = -2.4\arcmin$. Different black curves represent identified shock features in models with different $\Omega_{b}$ for $c_{s} = 30\kms$. The simulated velocities are projected with an inclination of $77^{\circ}$. \textit{Middle panel:} Similar to the left panel but for shock features in models with different $c_{s}$ for $\Omega_{b} = 18\freq$. \textit{Right panel:} Similar to the left panel but for shock features in a model projected with different inclinations. The model uses $\Omega_{b} = 30\freq$ and $c_{s} = 10\kms$.}
\label{fig:slit_4shock_feature_overview}
\end{figure*}

We present an example of how SMHD quantifies the similarity between two features by showing the best-fitting model in Figure~\ref{fig:KR241_bestfit_SD_pv}. This model was simulated using the KR241 potential, a low pattern speed of $\Omega_{b} = 18\freq$, and a high sound speed of $c_{s} = 34\kms$. The pattern speed of $\Omega_{b} = 18\freq$ places the corotation radius at $ R_{CR} \sim 14 \kpc$, resulting in $\mathcal{R}\equiv R_{CR}/a_{bar} \sim 3.5$. The gas morphology and kinematics of a similar model have been described in \citet{fen_etal_22}. For further details on the model results, we refer readers to \citet{fen_etal_22}. The bottom left panel of Figure~\ref{fig:KR241_bestfit_SD_pv} shows the face-on unprojected view of the model. The bar major axis (dashed line) is tilted by $\phi_{b} = 54.7^{\circ}$ from the disk major axis (the $X$-axis). The panel reveals a high-density gas ring in the nuclear region, which is connected to a pair of shocks on the leading side of the bar.  The top left panel shows the projected gas surface density with an inclination angle of $i = 77^{\circ}$. The shocks in the model match well with the observed positions of $\oiii$ (purple circles) and $\HI$ (green triangles) shock features. Large, medium, and small markers represent the Class I, Class II, and Class III shock features identified in \citet{fen_etal_22}, respectively. The first column of the right panels shows the gas PVDs of the model (black line) and the data of $\oiii$ (red points) and $\HI$ (blue plus signs). The model reproduces the main features of the observed gas kinematics with minor differences. The second column of the right panels also shows the zoom-in gas PVDs within the shock region, which is marked by the vertical dashed lines in the first column. The panel presents a more detailed comparison between shock features in models (black curve) and observations (red curve). The velocity features are presented in grids with a spacing of $\Delta p = 0.05\arcmin$ and $\Delta v = 8\kms$. Among the slits, S-4 exhibits a shock feature most similar to the $\oiii$ data, yielding a small SMHD of 1.5. Slits S-2 and S-3, on the other hand, show a larger difference in shock position and amplitude, resulting in a larger SMHD.

We also tested the results using the $\chi^{2}$ comparison method. Using the $\chi^{2}$ rather than the SMHD method does not change our main result. However, the $\chi^{2}$ results tend to show very high values for several reasons. First, our models are bisymmetric, but the observed $\oiii$ velocity field shows many asymmetries, such as shock features on the near side is farther from the disk major axis than those on the far side, and high-velocity features on the approaching side being farther from the disk minor axis than those on the receding side. Second, velocities of $\oiii$ are lower than those of our models near the center ($| X | < 6\arcmin$, $| Y | < 2.5\arcmin$, or $| X | < 1370 \pc$, $| Y | < 570 \pc$), which we discuss in \S\ref{sec:other_gas_feature}. Finally, $\chi^{2}$ is very sensitive to the position difference in the shock features. The strongest shock features produce large velocity jumps over $170 \kms$, so even a small positional shift of shock features can greatly increase $\chi^{2}$. In comparison, SMHD quantifies the overall similarity of the shapes of two shock features and ignores other anomalous features. Therefore, we prefer to use SMHD to characterize the comparison results.

\subsection{Convergence test on the spatial resolution}

Using higher spatial resolution in the simulations or later time period for the SMHD analysis have small effects on our results. We have made a comparison model using exactly the same parameters as the best-fitting model in Figure~\ref{fig:KR241_bestfit_SD_pv} but with 4096$\times$ 4096 cells (corresponding to a spatial resolution of $\Delta x \sim7.5\pc$). We find that the gas features in this model with higher resolution are very similar to the one presented in Figure~\ref{fig:KR241_bestfit_SD_pv}. The time-averaged SMHD for this gas model is $\sim4.5$, which is not much different from 3.9 for the best-fitting model. The gas flow becomes quasi-steady after two bar-rotation periods, so the properties used in the analysis do not sensitively depend on the chosen time period. Although transient fluctuations of velocity peaks in Figure~\ref{fig:Vrot_comparison} and zero-velocity lines can be spotted as time evolves, the main signatures that we described in \S~\ref{sec:other_gas_feature} are almost unchanged. We have also checked that changing the time cut in \S~\ref{sec:define_goodness_of_fit} from $T=800-1000\Myr$ to $1000-1200\Myr$ does not affect our main results in the SMHD analysis.

\section{Searching for best fitting models with a fixed inclination}
\label{sec:search_for_bestfit_model}

\subsection{Varying bar $\Omega_{b}$, $c_{s}$, and inclination}
\label{sec:vary_omegab_cs_I}

We first generate PVDs for S-2 using models that have different bar pattern speeds. The left panel of Figure~\ref{fig:slit_4shock_feature_overview} presents the PVDs for four models with varying bar pattern speeds, and compares them to the $\oiii$ shock feature (red curve). All four models use the KR241 potential and have a sound speed of $c_{s} = 30 \kms$. We use different line styles to represent the models with different bar pattern speeds. The solid curve corresponds to a model that approximately matches the red curve with $\Omega_{b}=18\freq$. As the bar pattern speed $\Omega_{b}$ increases from $14\freq$ to $30\freq$, the shock position moves closer to the bar major axis from $Y \sim -3.7\arcmin $ to $Y \sim -1.9 \arcmin$ (from $Y \sim -840\pc$ to $Y \sim -430\pc$). The figure shows that $\Omega_{b}$ primarily affects the shock positions.

\begin{figure*}[t!]
\includegraphics[width=\textwidth]{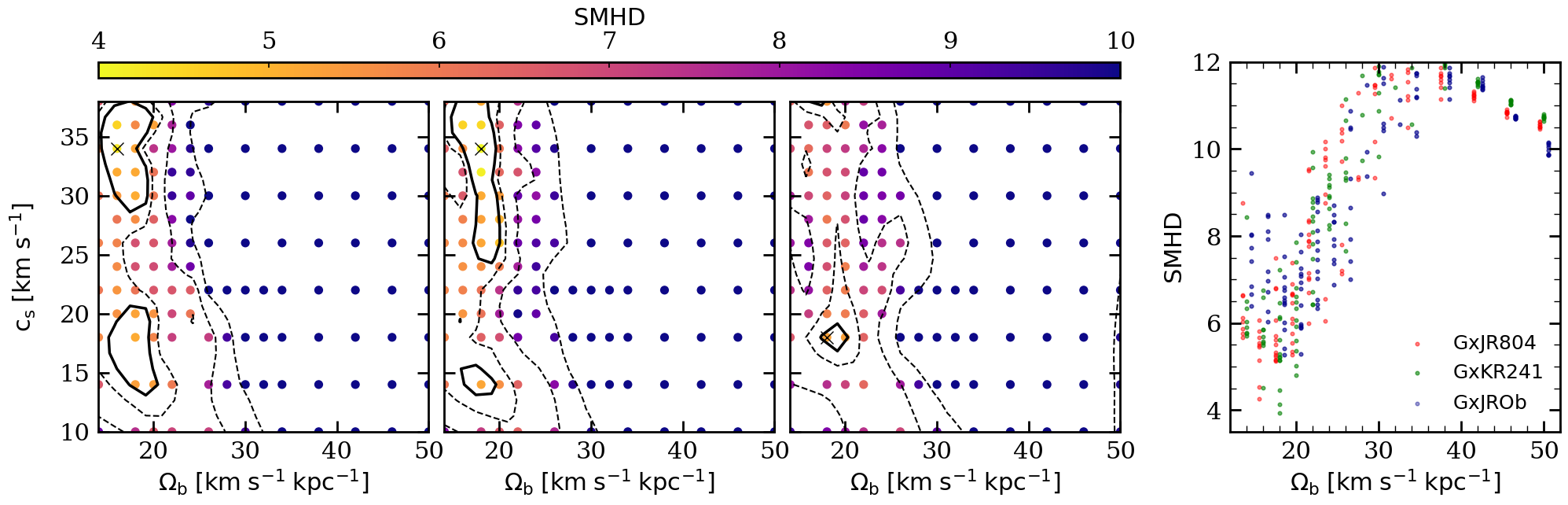}
\caption{\textit{Left panels:} Goodness of fittings of models in GxJR804 (left), GxKR241 (middle), and GxJROb (right) gas model groups. All models are projected with an inclination of $77^{\circ}$ and a bar angle $\phi_b$ of $55^{\circ}$. The points represent the positions of the models, color-coded with SMHD. Contours indicate the distribution of SMHD on the $\Omega_{b}$-$c_{s}$ parameter space. The contour levels are 5.5, 6.2, 8.0, and 10.0 from inside out. There are in total 127 models in each panel (see Table~\ref{tab:criteria_shock_feature}). The black crosses mark the best-fitting models in each panel. \textit{Right panel:} A marginalized plot showing SMHD as functions of $\Omega_{b}$. Red, green, and blue points represent the models in GxJR804, GxKR241, and GxJROb gas model groups, respectively. A local minimum of SMHD is clear to see at $\Omega_{b}=18 \pm 2 \freq$.}
\label{fig:overview_SMHD_I77}
\end{figure*}

The middle panel in Figure~\ref{fig:slit_4shock_feature_overview} shows the same context as the left panel, but for models with varying sound speeds. All models use a bar pattern speed of $\Omega_{b} = 18 \freq$. Previous gas simulations have found that shock features are sensitive to the sound speed \citep{kim_etal_12b, li_etal_15}, with positions shifting closer to the bar major axis as the sound speed increases. We indeed find that the shocks become more curved and away from the bar major axis ($Y = 0\arcmin$) as the sound speed decreases, and the velocity jumps on PVDs are closely related to the shape and positions of shocks. As the sound speed decreases from $c_s \sim 30\kms$ to $\sim 15 \kms$, the velocity jumps become much smaller. At a high sound speed of $c_s = 38\kms$,  however, the shock features become too far from the observed ones to give good fits. Although sound speed affects the shock positions as well, the shock positions are mainly determined by the bar pattern speeds. 

As noted in \citet{fen_etal_22}, a gas model with a smaller inclination of $i = 67^{\circ}$ and a pattern speed of $\Omega_{b} = 33 \freq$ can also approximately reproduce the observed shock features. Decreasing the inclination helps to project the shock features away from the disk major axis, resulting in better model fits for high pattern speeds.

If we consider the distance between shock features and the disk major axis to be $y_{shock}$ in the face-on view, then after projection, the distance becomes $Y_{shock} = y_{shock} \times \cos(i)$. For a new inclination of $i'$, the new distance would be $Y_{shock}' = y_{shock} \times \cos(i')$. The ratio between the two distances is $\frac{Y_{shock}'}{Y_{shock}} = \frac{\cos(i')}{\cos(i)}$. If we decrease the inclination from $77^{\circ}$ to $67^{\circ}$, the distance will be increased to $\cos(67^{\circ})/\cos(77^{\circ}) = 1.74$ times its original number. Such a ratio is sufficient to shift a shock feature close to the bar major axis to the positions of $\oiii$ shock features.

The right panel in Figure~\ref{fig:slit_4shock_feature_overview} demonstrates the effects of different inclinations on the shock positions. The red curve has the same meaning as in the left panels, and the different styles of black curves represent the velocities of a model projected with different inclinations. The model uses a bar pattern speed of $\Omega_{b} = 30 \freq$ and a sound speed of $c_{s} = 10 \kms$. The shock features are close to the bar major axis with such a pattern speed, producing velocity jumps at $Y \sim -2.5\arcmin$ ($-570\pc$) after projection of $i=77^{\circ}$. Decreasing the inclination from $77^{\circ}$ to $71^{\circ}$ significantly improves the fit of shock features by shifting the shock positions to $Y \sim -3.6\arcmin$ ($-820\pc$). However, it should be noted that changing inclinations do not affect the amplitude of shock features much.

\begin{figure}[t!]
\includegraphics[width=\columnwidth]{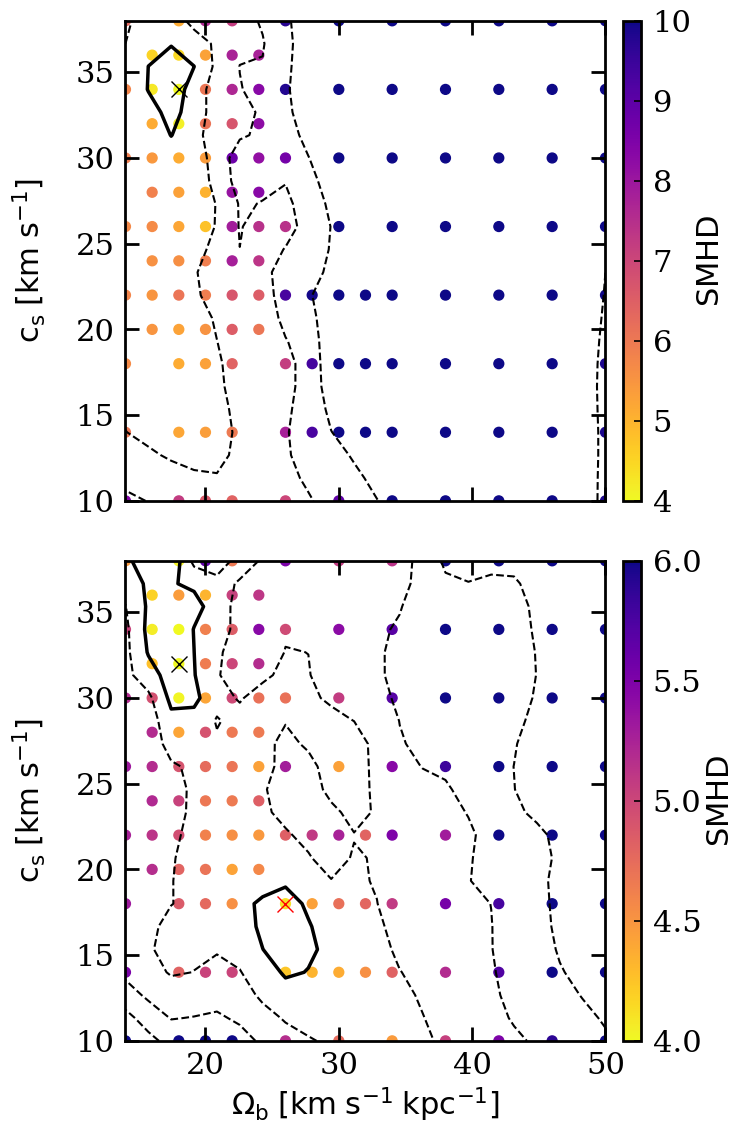}
\centering
\caption{\textit{Upper panel}: Similar to the left panels in Figure~\ref{fig:overview_SMHD_I77} but for colors representing the lowest SMHD of the three sets of models with different potentials at $(\Omega_{b},\; c_{s})$ position. \textit{Lower panel}: Similar to the upper panel but for each model the inclinations are allowed to change from $69^{\circ}$ to $77^{\circ}$ with a spacing of $2^{\circ}$ and the bar angles to change from $45^{\circ}$ to $65^{\circ}$ with a spacing of $5^{\circ}$. For each point models with three different potentials are projected with five different inclinations and five different bar angles, resulting in a total of $3\times5\times5 = 75$ combinations. The points are colored by the lowest number of SMHDs in the above combinations. Contours indicate the distribution of SMHD on the parameter space, with levels at 4.3, 4.8, 5.6, and 6.4. Two local minimums of SMHD with $(\Omega_{b},\; c_{s}) = (18\freq, 32\kms$) and ($26\freq, 18\kms$) are marked by the black and red crosses, respectively.}
\label{fig:overview_SMHD_changeI}
\end{figure}

\begin{figure}[t!]
\includegraphics[width=\columnwidth]{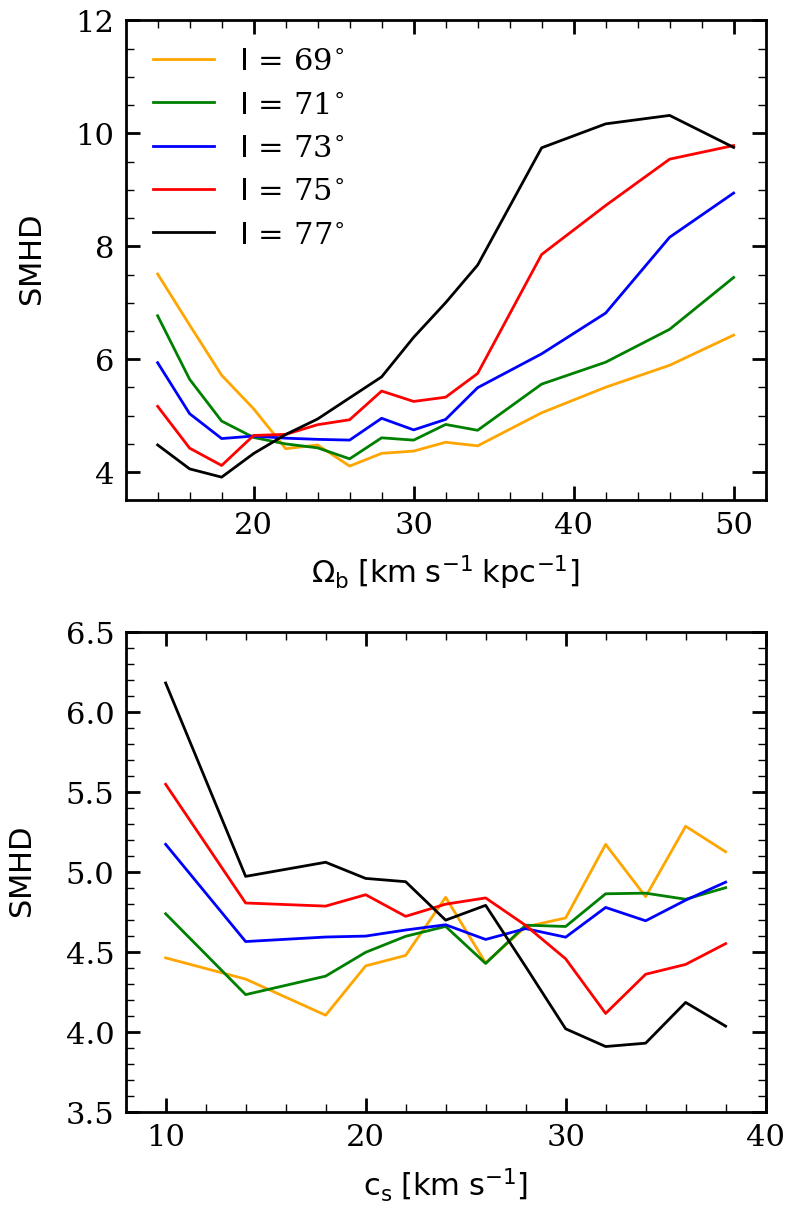}
\caption{\textit{Upper panel}: Curves showing the minimum SMHD as functions of $\Omega_{b}$ with different inclinations of the gas disk. \textit{Lower panel}: Similar to the upper panel but as functions of $c_{s}$.}
\label{fig:SMHD_curve_omegab_cs_changeI}
\end{figure}

\begin{figure*}[t!]
\includegraphics[width=\textwidth]{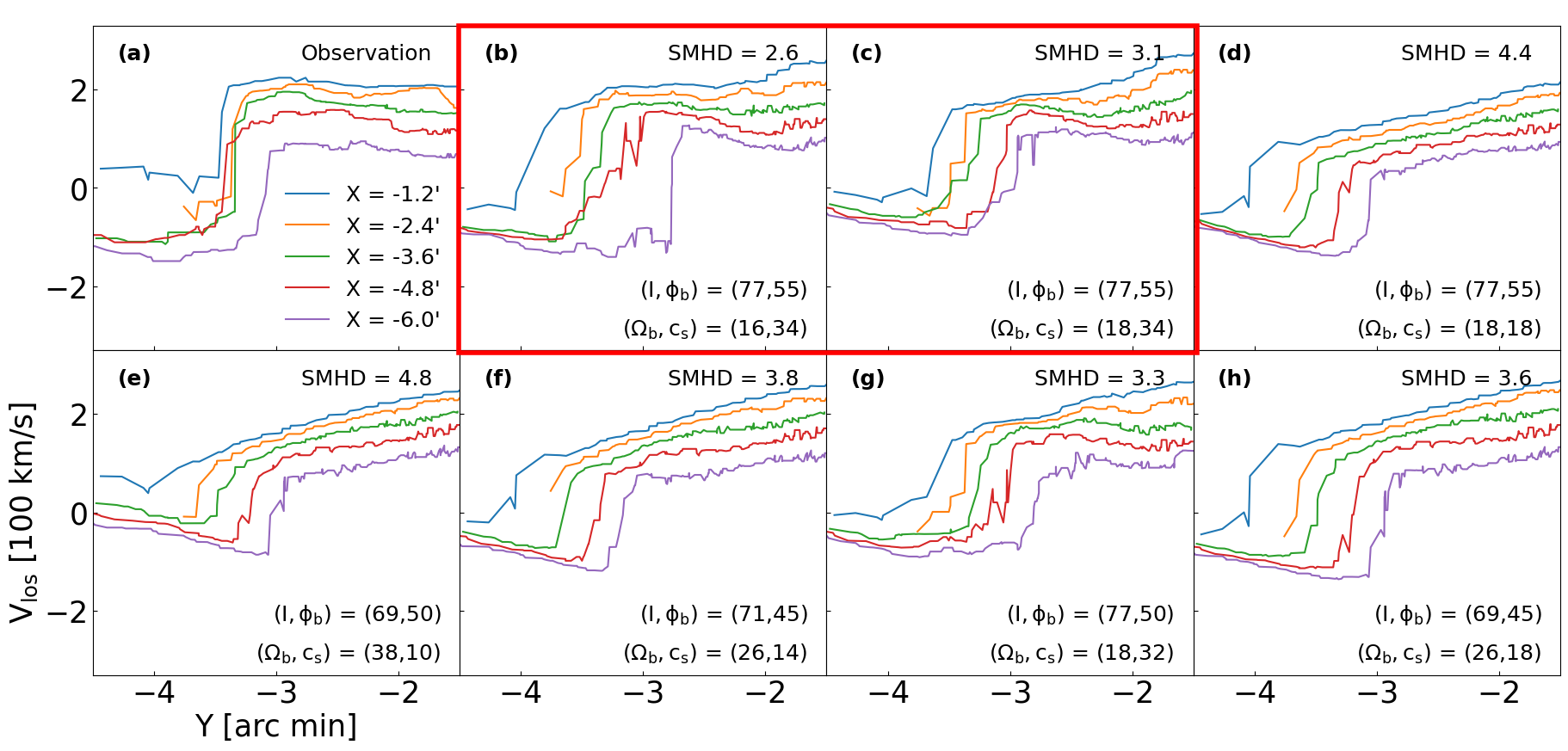}
\caption{\textit{First panel}: Identified shock features of $\oiii$ in five pseudo-slits perpendicular to the disk major axis of M31. The color represents the $X$ positions of the pseudo-slits. \textit{Other panels}: Same context as the first panel but for seven representative models with relatively small SMHD, with their parameters shown in the bottom right corner and SMHD shown in the top right corner. The red box highlights the two best-fitting models using the JR804 potential (b) and the KR241 potential (c).} 
\label{fig:SMHD_comparison_7models}
\end{figure*}

\subsection{Overall comparison}
\label{sec:SMHD_overall_comparison}
We present the comparison results of SMHD within all five pseudo-slits S-1 to S-5 in this section. For the idealization of our models, we do not intend to make a perfect fit to all $\oiii$ shock features, but to use them to give tight constraints on the bar parameters. We calculate SMHD for a series of models in the $(\Omega_{b},\;c_{s})$ parameter space with different m2m potentials (see \S~\ref{sec:grav_potential}). The left panels in Figure~\ref{fig:overview_SMHD_I77} present the distributions of SMHD in GxJR804, GxKR241, and GxJROb gas model groups in the first, second, and third columns, respectively. The points represent the positions of the models, color-coded with SMHD. Black crosses indicate the best-fitting models in each panel. The first and second columns show quite similar distributions of SMHDs. Models with low pattern speeds of $\Omega_{b} \leq 20 \freq$ fit the shock features much better than those with high pattern speeds. Although the overall distributions of SMHD are less sensitive to the change of sound speeds, a local minimum of SMHD can be recognized at a high sound speed of $c_{s} = 34 \kms$. The similarity of the SMHD results in the GxJR804 and GxKR241 gas model groups indicates that a $20\%$ difference in dark matter halo profile does not affect the main result. To investigate whether using m2m potentials with different $\Omega_b$ affect our main result, we construct the GxJROb gas model group and show its SMHD results in the third column of Figure~\ref{fig:overview_SMHD_I77}. The overall pattern of SMHD is similar to the left panels, with slightly increased SMHD values, particularly within the range of $\Omega_{b} = 16-20 \freq, c_{s} = 26-34 \kms$. The right panel of Figure~\ref{fig:overview_SMHD_I77} presents a marginalized plot showing the SMHD of all models as functions of pattern speeds, which indicates a clear minimum of SMHD at $\Omega_{b} = 18\pm2 \freq$. Red, green, and blue points represent the models in GxJR804, GxKR241, and GxJROb gas model groups, respectively. As $\Omega_{b}$ increases from $20\freq$ to $38\freq$, the SMHD increases significantly from $6\pm2$ to $12\pm1$.

To account for the effects of varying inclinations and bar angles, we projected each model on the left panels of Figure~\ref{fig:overview_SMHD_I77} with different inclinations ($i$) in the range of $67-77^{\circ}$ and different bar angles ($\phi_{b}$) in the range of $45-65^{\circ}$ with spacings of $2^{\circ}$ and $5^{\circ}$, respectively. Considering the three groups of potentials in Figure~\ref{fig:overview_SMHD_I77}, this resulted in a total of $5\times5\times3 = 75$ combinations for each point on the parameter space. The upper and lower panels of Figure~\ref{fig:overview_SMHD_changeI} show the goodness of fittings for models projected with (a) the fiducial inclination of $i = 77^{\circ}$ and bar angle of $\phi_{b} = 54.7^{\circ}$, and (b) different inclinations and bar angles, respectively. In the upper panel (case a), the color of each point in ($\Omega_{b}$, $c_{s}$) parameter space represents the lowest number of SMHDs among gas models in Figure~\ref{fig:overview_SMHD_I77}. In the lower panel (case b), the points are color-coded by the lowest number of SMHDs among the total 75 combinations described above.
The best-fitting model shown in Figure~\ref{fig:KR241_bestfit_SD_pv} corresponds to the black cross in the upper panel of Figure~\ref{fig:overview_SMHD_changeI}. In the upper panel of Figure~\ref{fig:overview_SMHD_changeI}, models with larger pattern speeds of $\Omega_{b} \geq 30\freq$ show SMHDs greater than 10, which are much larger than those of the models with lower pattern speeds of $\Omega_{b} \sim 20 \freq$. In the lower panel of Figure~\ref{fig:overview_SMHD_changeI}, the SMHDs of models with larger pattern speeds of $\Omega_{b} > 34\freq$ reduce significantly from over 10 to $\sim 5$. For models with $\Omega_{b} < 30 \freq$, two local minimums of SMHD appear at ($\Omega_{b}$, $c_{s}$) $=$ (18, 32) and (26, 18), indicated by the black and red crosses, respectively. The former is a variant of the best-fitting model in Figure~\ref{fig:KR241_bestfit_SD_pv}, and the latter is a representative of the models with smaller inclinations. 

In Figure~\ref{fig:SMHD_curve_omegab_cs_changeI}, we present curves that quantify the effects of inclinations on the SMHD results. The upper and lower panels show SMHD as functions of $\Omega_{b}$ and $c_{s}$, respectively. Note that the black curves are not the same as the lower envelope of the marginalized plot in Figure~\ref{fig:overview_SMHD_I77} because we allow $\phi_b$ to change in the range of $45-65^{\circ}$. The colored curves show a similar context but for models projected with different inclinations of $77^{\circ}$, $75^{\circ}$, $73^{\circ}$, $71^{\circ}$, and $69^{\circ}$, which are shown in black, red, blue, green, and yellow, respectively. The upper panel demonstrates a clear trend that larger and smaller inclinations prefer models with lower and higher bar pattern speeds, respectively. As the colors of the curves change from black to yellow, the SMHD of models with $\Omega_{b} \leq 18 \freq$ increase from 4 to 6-8, while the SMHD of models with $\Omega_{b} \geq 45 \freq$ decrease from 10 to 6. In addition, the minimum of SMHD shifts from $\Omega_{b} = 18 \freq$ to $26 \freq$ as the inclination decreases from $77^{\circ}$ to $69^{\circ}$. The lower panel illustrates that larger and smaller inclinations prefer models with higher and lower sound speeds, respectively. For instance, at a large inclination of $i = 77^{\circ}$, it requires a low pattern speed of $\Omega_{b}\sim 18 \freq$ to produce shock positions as far from the disk major axis as those in $\oiii$. In this case, the sound speed should be relatively large, otherwise, the shock positions will be even farther than those in $\oiii$, as discussed in \S~\ref{sec:vary_omegab_cs_I}. At a small inclination of $i = 69^{\circ}$, a pattern speed of $\Omega_{b}\sim 26 \freq$ produces shock positions similar to observations, and a relatively low sound speed helps fine-tune the fittings. 

\begin{figure*}[t!]
\includegraphics[width=\textwidth]{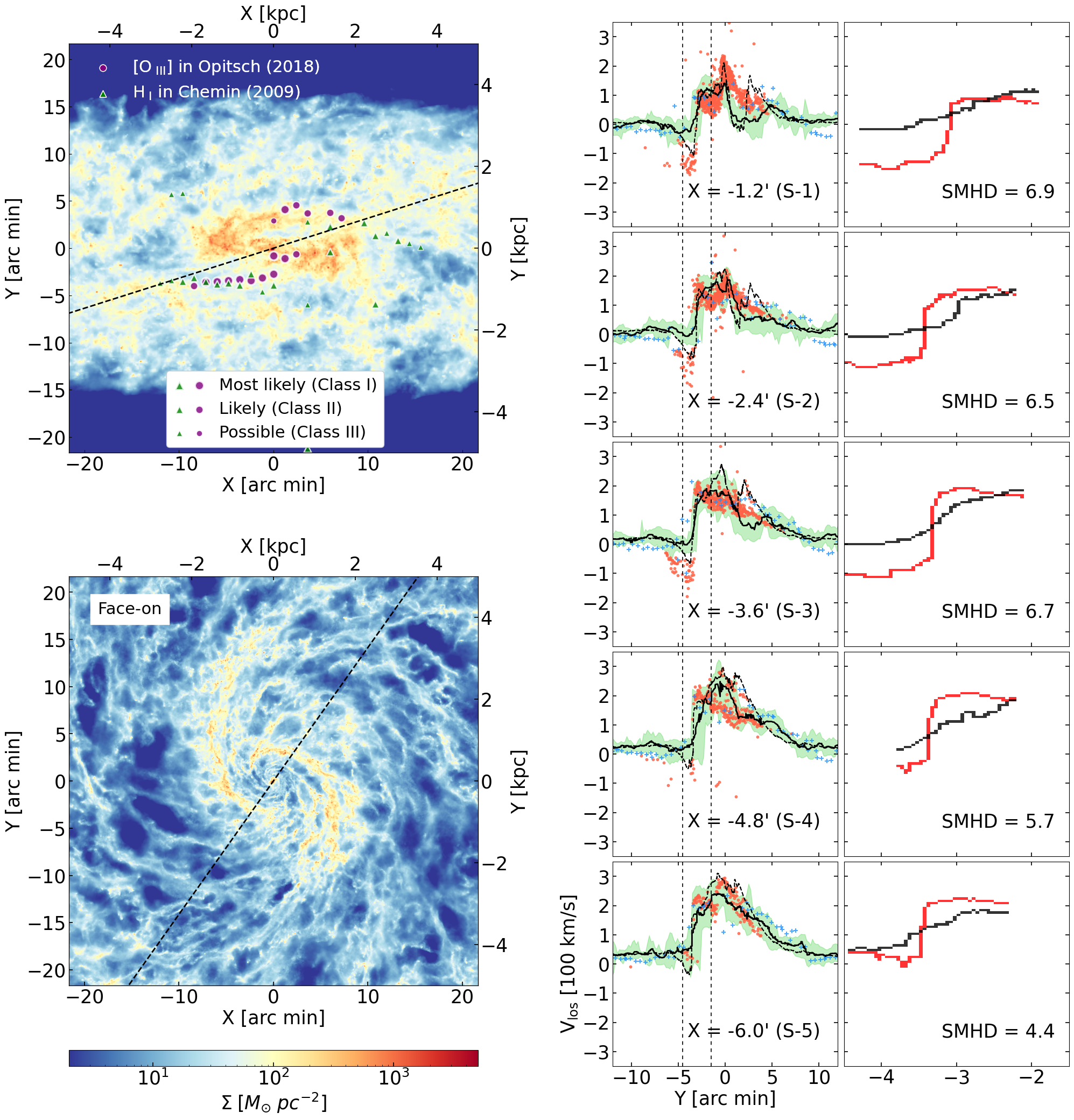}
\caption{Similar to Figure~\ref{fig:KR241_bestfit_SD_pv} but for a SMUGGLE model with $\Omega_{b} = 18 \freq$. The snapshot is taken at $T = 840\Myr$. The green shaded regions in the right panels show the 10-90 percentile range of the PVDs of all gas cells. The black curves represent the boxcar smoothed result of the density-weighted line-of-sight velocities. The dashed curves represent the PVDs of the best-fitting model in Figure~\ref{fig:KR241_bestfit_SD_pv}. The second column of the right panels show a zoom-in view of PVDs in the shock region, which is indicated by the vertical dashed lines in the first column.}
\label{fig:Ob18_smuggle_fit_SD_pv}
\end{figure*}

\subsection{Shock features in several representative models}
In Figure~\ref{fig:SMHD_comparison_7models}, shock features in several representative models with relatively low SMHD are presented, along with their SMHD and model parameters displayed in the top right and bottom right corners, respectively. Panel (a) shows the shock features in $\oiii$, with colors indicating the positions of the pseudo-slits on the disk major axis. Panels (b), (c), and (d) present the shock features of the best-fitting models in Figure~\ref{fig:overview_SMHD_I77}, which use the JR804, KR241, and JR644 potentials, respectively. All of these models are projected with an inclination of $i = 77^{\circ}$ and a bar angle of $\phi_{b} = 55^{\circ}$. We used the snapshots that exhibit shock features most similar to those in the $\oiii$ data, resulting in smaller SMHD in Figure~\ref{fig:SMHD_comparison_7models} compared to the time-averaged ones in Figures~\ref{fig:overview_SMHD_I77}. As discussed in \S4, although reducing the inclination of the gas disk helps improve the fits for models with higher pattern speeds, it mainly affects the shock positions and does not improve the shock amplitude. Such effects are reflected in the large SMHD of 4.8 in panel (e), where we present a model with JR804 potential and the fiducial bar pattern speed of $\Omega_{b} = 38\freq$. Panels (f), (g), and (h) show the same context as (b), (c), and (d) but for the best-fitting models with different inclinations and bar angles. Among these models panels (b) and (c) present shock features most similar to those in $\oiii$, with SMHD less than 3.2. 

\section{More sophisticated gas models}
\label{sec:smuggle_model}

\subsection{Initial conditions}
Gas in late-type galaxies is multiphase, which motivates us to test our dynamical models with more realistic gas physics with explicit cooling/heat, star formation, and stellar feedback. The Stars and MUltiphase Gas in GaLaxiEs (SMUGGLE) module \citep{mar_etal_19} in the moving-mesh hydrodynamic code AREPO \citep{volker_10} provides an ideal tool to investigate how realistic gas physics affects the observables mentioned in the above section. The SMUGGLE module includes physical mechanisms of radiative cooling, and heating, star formation, and stellar feedback. Star particles in the simulation are formed from cold dense gas in cells with density above a threshold of $n_{\rm th} = 100 \cm^{-3}$. The star formation rate $\dot{M}_{\star}$ is calculated based on a given star formation efficiency $\epsilon_{\rm ff}$ with a relation $\dot{M}_{\star} = \epsilon_{\rm ff} M_{\rm gas}/\tau_{\rm ff}$, here $\tau_{\rm ff}$ is the free-fall time-scale of a gas cell. During the star formation process, each star particle represents a stellar population that follows the Chabrier initial mass function \citep{chabri_03}. 
The total energy deposited in a single supernova event is determined by $E_{\rm SN} = f_{\rm SN}E_{\rm 51}$, here $f_{SN}=1$ represents the feedback efficiency and $E_{\rm 51}=10^{51} \rm erg$. The mass, momentum, and energy injection from stellar winds and supernova events, as well as the impacts of photoionization and radiation pressure are all taken into account. We refer readers to \citet{mar_etal_19} for more details about the settings in the feedback module. The SMUGGLE framework has been shown to produce a realistic multiphase ISM \citep{mar_etal_19}, dense ISM and star cluster properties in simulated late-type and merging galaxies \citep{lihui_etal_20, lihui_etal_22}, and constant-density cores in idealized dwarf galaxies \citep{jahn_etal_21}. To study the detailed feature of gas response in the designed gravitational potential, we use a mass resolution of $\sim 1.4\times10^{3} \Msun$ for both gas and stars. The gas cells have a minimum gravitational softening length of $3.6\pc$, same as that of star particles. The gas surface density in the 3D models follows the profile: 

\begin{equation}
\rho_{\rm gas}(R,z) = \dfrac{\Sigma_{0}}{2z_{\rm gas}}\exp(-R/R_{d}) \sech^{2}(z/z_{\rm gas}). 
\end{equation}

Here we use $z_{\rm gas} = 0.26 \kpc$ as constrained by $\HI$ observation in \citet{bra_etal_09}. The initial settings of gas velocities are similar to those described in \S~\ref{sec:initial_setting}. We first allow the gas to evolve with self-gravity under an azimuthally averaged gravitational potential and following an adiabatic equation of state for 400$\Myr$. Then we activate the SMUGGLE module using a star formation efficiency of 0.01 and a feedback efficiency of 1.0. During this process, the gas rapidly cools down in regions with large gas density, resulting in a vertically thinner gas disk within several $\Myr$. The above procedure enables us to generate a gas disk that is in quasi-equilibrium. At the same time, we slowly ramp up the non-axisymmetric gravitational potential until the barred potential is fully reached over 100 $\Myr$, same as in \S~\ref{sec:initial_setting}. Our input parameters of SMUGGLE are the same as the fiducial settings described in \citet{mar_etal_19}, except for the number of nearest effective neighbors $N_{ngb}$. \citet{bea_etal_22} found that the fiducial number of $N_{ngb} = 64$ leads to inefficient diffusion of the photo-ionization feedback energy. Therefore, they recommend a higher value of $N_{ngb} = 512$ to better quantify the process of photo-ionization feedback, which we adopt in our high-resolution models. 


\subsection{Similarity and difference between SMUGGLE and isothermal models}

The Athena++ 2D simulations show a preferred $\Omega_{b}$ in the range of 18 $\pm$ 2$\freq$ based on the systematic exploration of SMHD (see \S~\ref{sec:SMHD_overall_comparison}). We conducted tests with a SMUGGLE model that uses $\Omega_{b} = 18 \freq$. Appendix~\ref{appendix:eos_comparison} tests the difference between the more sophisticated model with SMUGGLE turned on in the AREPO code and the 2D/3D isothermal models with Athena++ code. Our results indicate that the overall gas patterns of the SMUGGLE runs are similar to those in the isothermal runs with sound speeds of $c_{s}\sim 15\kms$. According to \S~\ref{sec:vary_omegab_cs_I}, gas models with a sound speed of $c_{s}\sim 15\kms$ produce shocks away from the bar major axis, resulting in shock positions lower than those in $\oiii$. Therefore, to better fit the $\oiii$ shock features, a more turbulent gas model is necessary. To achieve this, we increase the star formation efficiency $\epsilon$ from 0.01 to 0.03 and the stellar feedback efficiency $f_{\rm SN}$ from 1.0 to 2.0 to enhance the level of gas heating. These numbers are chosen empirically to produce shock features at positions similar to those in $\oiii$.

Figure~\ref{fig:Ob18_smuggle_fit_SD_pv} displays the gas surface density and kinematics of the SMUGGLE model, with markers and colors having the same meanings as those in Figure~\ref{fig:KR241_bestfit_SD_pv}. Although the supernova explosions create many holes that compress the gas near their edges and complicate the overall gas pattern, the gas substructures produced by the bar can still be identified well. The face-on view (bottom left panel) shows a high-density nuclear ring connected with a pair of shocks, which are located on the leading side of the bar and extend roughly to the end of the bar. It is worth noting that the best-fitting isothermal model in Figure~\ref{fig:KR241_bestfit_SD_pv} uses an effective sound speed of $c_{s} = 34 \kms$, representing a very turbulent gas pattern. As discussed in \S~\ref{sec:vary_omegab_cs_I}, the level of gas turbulence affects the gas morphology; therefore, we do not expect the bar-driven substructures in Figures~\ref{fig:KR241_bestfit_SD_pv} and \ref{fig:Ob18_smuggle_fit_SD_pv} to have the same sizes and positions. When projected with an inclination of $77^{\circ}$ (top left panel), a pair of gas filaments appear at positions similar to those of the $\oiii$ shock features. The gas disk presents an overall higher surface density, which is due to projection effects.

Figure~\ref{fig:Ob18_smuggle_fit_SD_pv} displays more complicated gas PVDs than Figure~\ref{fig:KR241_bestfit_SD_pv}. Supernova explosions from the stellar clusters produce small velocity jumps near low-density cavities, while projection effects of 3D models broaden gas features on PVDs. The green shaded areas on PVDs represent gas features within $10-90$ percent levels of those of the entire gas cells. We use the density-weighted average of the green shaded areas (black solid curves) to describe their trend on PVDs and plot the gas PVDs of the best-fitting isothermal model (black dashed curves) for comparison. The overall gas features indicated by the green shaded regions are similar to those in the $\oiii$ data, with slight differences in shock amplitude. The second column shows comparisons of shock features between the SMUGGLE model (black curves, same as those in the first column) and the $\oiii$ data (red curves). The black curves present smoother shock features with smaller amplitudes compared to the red curves, leading to an average SMHD of $\sim 6$. Cooling effects are prominent near high-density shock regions, resulting in an average gas temperature of around 10000 $K$. The temperature of $10000\;K$ corresponds to a gas sound speed of $\sim 10 \kms$, which together with the velocity dispersion inside the bar region of $\sigma \sim 40 \kms$ produces a local effective sound speed $c_s$ of $\sqrt{40^2+10^2} \sim 42\kms$ near the shock region. However, it should be noted that the effective sound speed of $\sim 30 \kms$ in the best-fitting isothermal model represents a global property of gas. The best-fitting isothermal model produces velocity dispersion as high as 100 $\kms$ near the shock region, much higher than the one in the SMUGGLE model. It is reasonable that the shock velocity jumps in the less turbulent SMUGGLE model are smaller than the $\oiii$ data, as we have presented in Figure~\ref{fig:slit_4shock_feature_overview}. 


\section{Gas models with disk following varying inclinations}
\label{sec:model_varying_inclination}
Although the best-fitting model in \S~\ref{sec:search_for_bestfit_model} reproduces the $\oiii$ shock features reasonably well, it is not fully consistent with other observed central gas features (see \S~\ref{sec:other_gas_feature}). In addition, the pattern speed of the best-fitting gas model $\Omega_b = 18\freq$ is much lower than $\Omega_b =40\pm5\freq$ in the stellar-dynamical m2m model (see more details in \S~\ref{sec:omegab_discrepancy}). If the gas disk is co-planar with the stellar disk, our gas simulations of M31 indicate that the bar should rotate at a pattern speed of approximately 18$\freq$, as demonstrated in \S~\ref{sec:search_for_bestfit_model}. However, a minor merger occurring near the center of M31, as proposed by \citet{blo_etal_06}, may tilt the inner gas disk without significantly affecting the gravitational potential of the stellar bulge and disk. It is possible that the bar in M31 could rotate at a higher pattern speed, and a head-on collision subsequently decreases the inclination of the inner gas disk, moving the shocks to positions of $\oiii$ shock features. In this section, we try to investigate if a model with a higher pattern speed of $\Omega_b = 38\freq$ and varying gas disk inclination angles can fit observed shock positions. We refer readers to \S~\ref{sec:other_gas_feature} for more details of the central gas features in this model. Note that we do not intend to find a perfect match of shock features using this model, but aim to test if a tilted inner gas disk helps to alleviate the discrepancy for the bar pattern speed measurement between stellar and gas dynamical models. Further research on interactions between M31 and its satellites is required to investigate this possibility.

\begin{figure}[t!]
\includegraphics[width=\columnwidth]{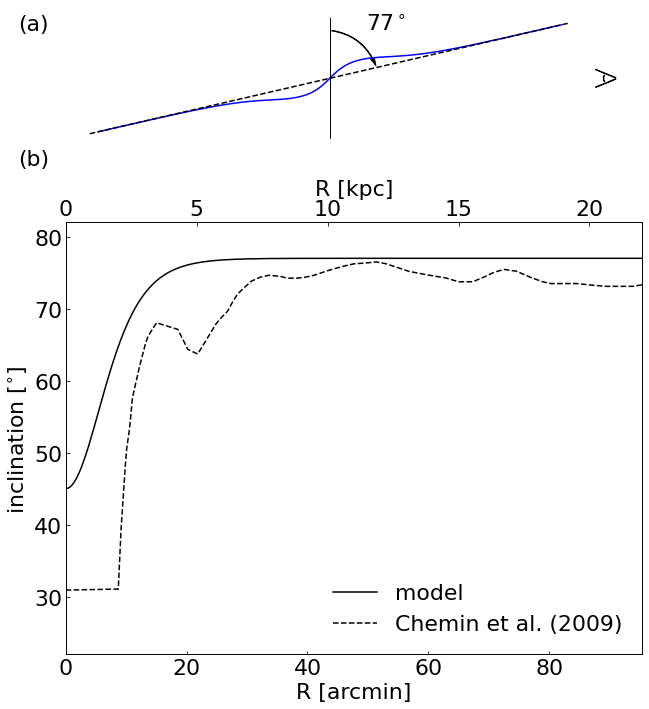}
\caption{\textit{Upper panel}: a schematic view of the gas disk in the model with varying inclination. \textit{Lower panel}: inclinations of the gas disk as functions of radius. Solid line represents the distribution of inclination in the gas model. Dashed line indicates the result of tilted ring fitting on $\HI$ data in \citet{che_etal_09}. Note that in our gas model the line-of-nodes is along the disk major axis at $PA_{disk} = 38^{\circ}$.}
\label{fig:Ob38c18_i_scheme}
\end{figure}

\begin{figure*}[t!]
\includegraphics[width=\textwidth]{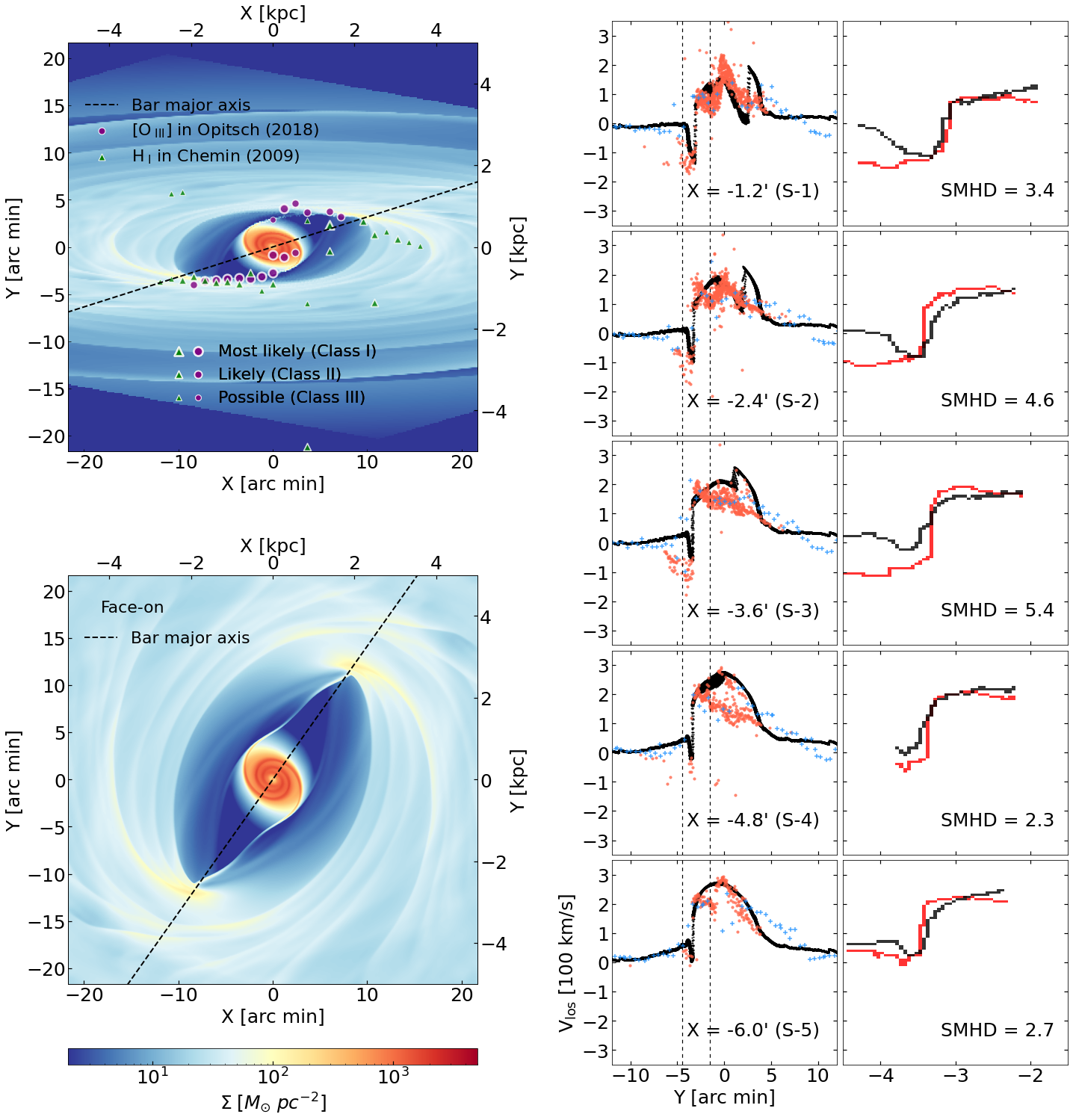}
\caption{Similar to Figure~\ref{fig:KR241_bestfit_SD_pv} but for a model with $\Omega_{b} = 38 \freq$, $c_s = 18\kms$ and varying inclinations. The snapshot is taken at $T = 850\Myr$.}
\label{fig:Ob38c18_varyingi_SD_pv}
\end{figure*}

The flux density of ionized gas in \citet{opi_etal_18} shows "face-on" spiral patterns that are tilted from the stellar disk in the inner $200\arcsec \times 200\arcsec$ ($760\pc \times 760\pc$) of M31. The gas morphology is similar to that observed in $\rm H\alpha+[N_{II}]$ and $\rm [O_{III}]$ \citep{jac_etal_85, cia_etal_88}. The $\HI$ survey conducted by \citet{che_etal_09} also noted a difference in inclinations between the inner and outer gas disks. By fitting the kinematics of $\HI$ with a group of tilted rings, they found that the inclination of the $\HI$ disk inside $R = 20 \arcmin$ ($\sim 4.6\kpc$) is less than $70^{\circ}$, which is lower than the average inclination of the outer $\HI$ disk of approximately $74^{\circ}$. The inclination even decreases considerably as the radius approaches the center. \citet{mel_com_11} found that the line-of-sight velocities of CO show two components in the central $1.5\kpc \times 1.5\kpc$ region of M31. They attribute the main component to the rotation of the tilted inner disk, and the second component to the perturbation caused by a recent merger between M31 and M32 \citep{blo_etal_06}. 
Recently, \citet{tre_etal_20} constructed multi-phase gas simulations of the Milky Way with a barred potential and found that the accretion of gas inflow can tilt the CMZ by $\sim 1-5 ^{\circ}$. Although a tilt of $\sim 5^{\circ}$ is small compared to the inner ring in M31, the tilted CMZ is long-lasting in their simulations. We have used the tilted ring fitting method with {\tt rotcur} task \citep{begema_89} in the NEMO software \citep{teuben_95} to investigate if the lower central inclination inferred by \citet{che_etal_09} can be explained by the
non-circular velocity field of the gas flow in a barred potential. Our tests show that with an initial inclination of $i=77^{\circ}$, the fitted inclination appears nearly constant at the correct value for a model with $\Omega_b=38\freq$. Although a change of inclination by $\Delta i \sim 10^{\circ}$ appears in the bar region when pattern speed decreases to $\Omega_b=18\freq$, the fitted inclination returns to $i\sim77^{\circ}$ in the nuclear ring region. Therefore the tilted disk inferred by \citet{che_etal_09} is likely to be real.

We set up an inclination profile with a trend similar to the observed ones:

\begin{equation}
\label{eq:iprofile}
-A\sech(x/\sigma)^2+77,
\end{equation}

here $A=-32$ is a scaling factor and $\sigma=8.2$ determines the sharpness of the profile. The numbers are chosen to make the profile smoothly increase from $i=45^{\circ}$ at the center to $i=77^{\circ}$ in the outer region with $R>25\arcmin$ ($5.7\kpc$). In panel (a) of Figure~\ref{fig:Ob38c18_i_scheme} we show a schematic diagram of the corresponding gas disk construction of equation~\ref{eq:iprofile}. We compare the inclination profile (solid curve) with the tilted ring fitting result of $\HI$ velocity field in \citet{che_etal_09} (dashed curve) in panel (b) of Figure~\ref{fig:Ob38c18_i_scheme}. The panel indicates that the inclination of $\HI$ disk decreases as the position moves inwards to the center. We do not intend to match the dashed line because the $\HI$ data is too scarce inside $10 \arcmin$ ($\sim 2.3\kpc$) and the fitted inclinations are overall irregular. Instead we aim to create a smooth curve following a trend similar to the dashed curve. Inside the bar region ($R < 4\kpc$), the solid curve is slightly higher than the dashed one because it requires such larger inclinations to produce shock features at positions similar to those in the $\oiii$ data \footnote{We have tested a lower inclination profile that is closer to the dashed curve. Our results show that such a profile projects shock features to $| Y |$ distances larger than the observed $\oiii$ shock positions.}.

Figure~\ref{fig:Ob38c18_varyingi_SD_pv} presents the same context as Figure~\ref{fig:KR241_bestfit_SD_pv} but for the model with $\Omega_b=38\freq$, $c_s=18\kms$, and varying inclinations. In the face-on view, the model with a higher pattern speed of $\Omega_b=38\freq$ produces shock features much closer to the bar major axis. However, if the gas disk is allowed to follow a varying inclination as in Figure~\ref{fig:Ob38c18_i_scheme}, the shocks can be projected to the observed $\oiii$ shock positions. Near center, the nuclear ring is less inclined compared to the outer gas disk, producing a more "face-on" pattern inside inner $3\arcmin \times 3\arcmin$ ($\sim 680\pc \times 680\pc$). The right panels show that the shock features and overall gas PVDs in the model and data are similar, resulting in an average SMHD of $\sim 3.7$.

\section{Discussion}
\label{sec:discussion}


\subsection{Effects of a different bar angle}
\label{sec:different_phib}
The main observational signature for the low pattern speed in gas models is that the shock features are found at a large distance from the bar major axis. When the bar is more end-on, the shocks are expected to be found at larger distances from the disk major axis. However, in our tests increasing bar angles $\phi_b$ does not help improve the models with $\Omega_b\sim40\freq$. As $\phi_b$ increases, the shocks become more extended in $Y$ but less extended in $X$. The effects of shifting shocks to larger $Y$ distances are not significant until $\phi_b$ increases to $75^{\circ}$. The models cannot reproduce shock features at large $X$ distances from the center (e.g. the shock features at $X<-4.8'$) with such a large bar angle. However, it should be noted that if the bar in M31 is more end-on, maybe the intrinsic bar length would be longer and the $\mathcal{R}$ ratio would be less extreme.

\subsection{Other observed gas features}
\label{sec:other_gas_feature}

\begin{figure}[t!]
\includegraphics[width=\columnwidth]{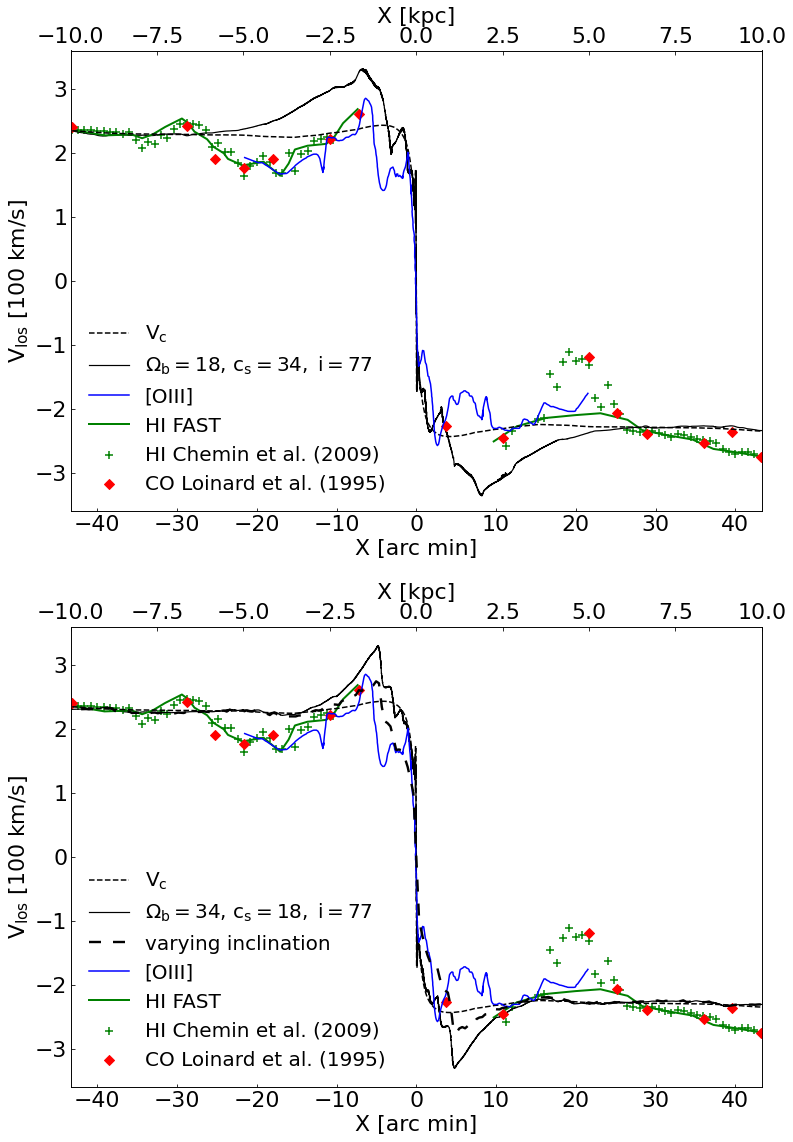}
\caption{\textit{Upper panel}: Gas velocities along the disk major axis of M31 in our best-fitting model with $i=77^{\circ}$ and the comparison with the observations. The dashed curve represents the projected circular rotation curve of the model. The black solid curve represents the gas line-of-sight velocities in the model at $T = 800\Myr$. The blue curve, green plus signs, and red diamonds indicate the data of $\oiii$, $\HI$, and $\rm CO$, respectively. The green curves represent the envelope of the HI PVD in the archival data of the FAST observation. \textit{lower panel}: same as the upper panel but for the model with $\Omega_b=38\freq$ and $c_s=18\kms$. The solid and long dashed lines represent the same model but projected with $i=77^{\circ}$ and varying inclinations, respectively.}
\label{fig:Vrot_comparison}
\end{figure}

\subsubsection{Velocity profile along the disk major axis}
The IFU observations by \citet{opi_etal_18} provide a comprehensive view of the gas kinematics in the bulge region of M31. In addition to the large-scale shock features, several high-velocity peaks of $\oiii$ are observed on the disk major axis, indicating non-circular motions in a non-axisymmetric potential. Such high-velocity peaks have also been observed in $\HI$ \citep{che_etal_09} and $\rm CO$ \citep{loi_etal_95}. Figure~\ref{fig:Vrot_comparison} presents the velocity distributions of gas along the disk major axis with different tracers of $\oiii$ (blue curve), $\HI$ (green plus signs), and $\rm CO$ (red diamonds). Note that multiple components are found in the $\HI$ emission lines \citep{che_etal_09}. In Figure~\ref{fig:Vrot_comparison}, we use the main component of $\HI$ that traces the velocities of the inner gas disk. The more recent $\HI$ survey by the Five-hundred-meter Aperture Spherical Radio Telescope \citep[FAST;][]{nan_etal_11,li_pan_16} has observed the $\HI$ structures in M31 with higher sensitivity. We refer the readers to Appendix~\ref{appendix:HI_obs} for the observation details. This FAST data are represented by the green curve, which is closely aligned with the main component of $\HI$ observed by \citet{che_etal_09}. We did not plot the green curve near the central region within $| X | < 10\arcmin$ ($\sim 2.3\kpc$) because the $\HI$ features there are quite faint. The FAST data also reveal gas features that have not been detected in previous $\HI$ observations. The green curve shows a hump feature with an amplitude of $\sim 40\kms$ at $X\sim 20\arcmin$ ($4.6\kpc$), while the plus signs present a bump feature with a large amplitude over $90\kms$ at the same position. \citet{che_etal_09} may have mistaken several faint features on the disk at $X\sim20'$ for the high-velocity cloud. The authors excluded them when obtaining the $\HI$ main component, so the plus signs show lower velocities compared to the green curve and appear as a bump feature. 
The observed velocities in $\oiii$, $\HI$, and $\rm CO$ are roughly similar, showing slightly lower velocities near $| X |\sim 3-6 \kpc$, and a flat part of $V_{flat}\sim250\kms$ in the outer region. The circular rotation curve in the model is presented by the black dashed curve. The solid curves in the upper and lower panels of Figure~\ref{fig:Vrot_comparison} represent the gas velocities in the best-fitting model with $\Omega_b=18\freq$, $c_s=34\kms$ and a comparison model with $\Omega_b=38\freq, c_s=18\kms$, respectively. Both models use an inclination of $i=77^{\circ}$. The long dashed curve in the lower panel indicates the same model as $\Omega_b=38\freq$, but allowing the gas disk to follow varying inclinations.

\begin{figure}[t!]
\includegraphics[width=\columnwidth]{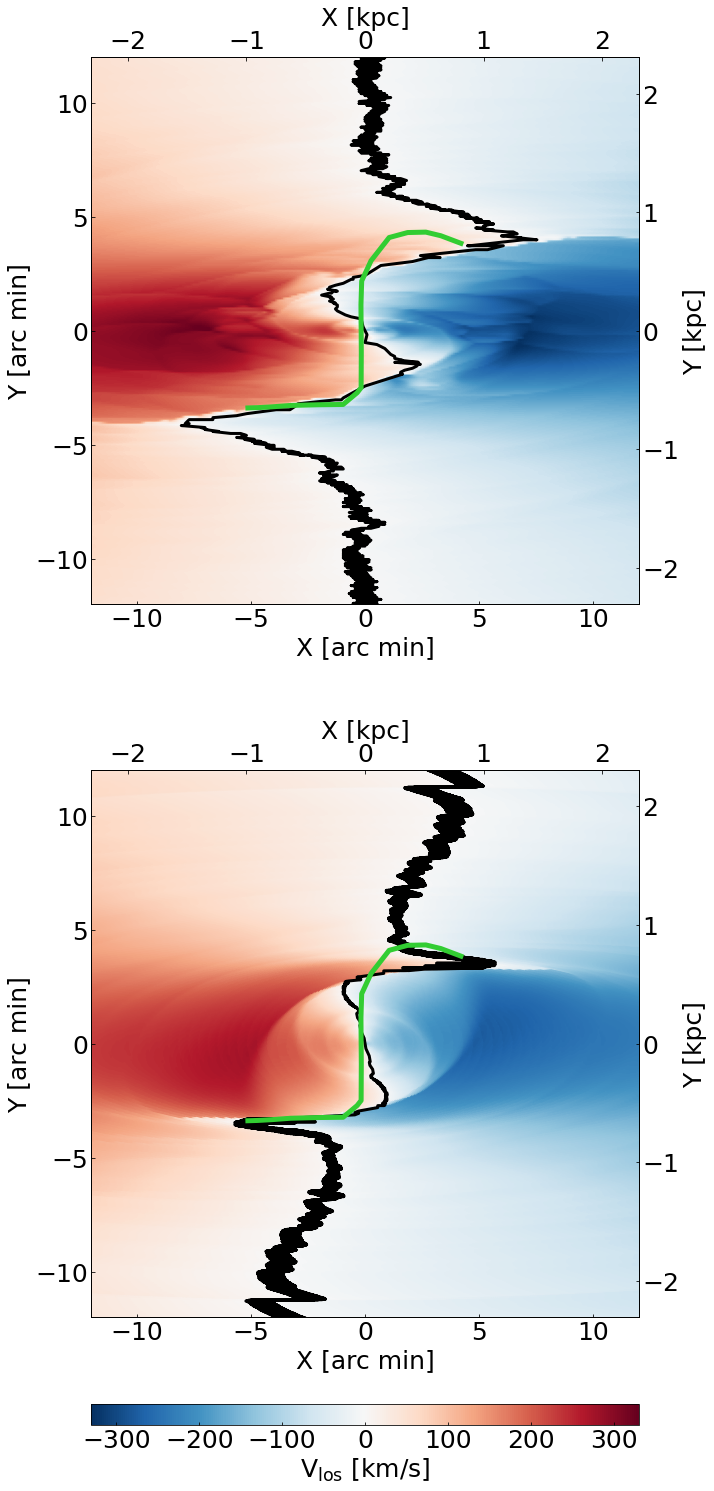}
\caption{\textit{Upper panel}: A map showing the velocity field of the best-fitting model with a fixed inclination $i=77^{\circ}$ in Figure~\ref{fig:KR241_bestfit_SD_pv}. Black area indicate regions with $| V_{los} | < 2.5\kms$ in the gas model and can be used to mimic the zero-velocity curve. Green curves indicate the zero-velocity curve in the $\oiii$ data. \text{Lower panel}: Same as the upper panel but for the model projected with varying inclinations in Figure~\ref{fig:Ob38c18_varyingi_SD_pv}.}
\label{fig:ZVL_comparison}
\end{figure}

In the upper panel of Figure~\ref{fig:Vrot_comparison}, our simulations show that gas follows nearly circular motions within the nuclear ring (radius $\sim 1 \kpc$ ), and is dominated by non-circular motions in regions where the $x_{1}$ orbits are present. Inward gas flow approaches the pericenter because it is near the edges of nuclear rings, producing high-velocity peaks of $V_{los}\sim330\kms$. The sharp inward decrease of velocities near the high-velocity peaks indicates a transition between the non-circular and the circular motions of gas. Near the center, the velocity peaks of $V_{los}\sim240\kms$ are caused by the change of angular momentum as gas crosses the nuclear spirals. Although the velocity features in the model show a shape similar to the observations, their values are much higher than those observed inside $R = 6\kpc$. We also plot the circular rotation curve $V_c$ (black dashed) for comparison. Note that the circular rotation curve of the model is also higher than most observed features inside 6 $\kpc$ due to the mass contribution from the classical bulge. In our tests, the differences between models (both $V_{los}$ and $V_c$) and observations are systematic and are not sensitive to the change in bar parameters. Improving the fittings requires either a smaller inclination of the central gas disk or a less massive bulge/bar. However, as discussed in \S~\ref{sys_uncertainty}, the dynamical mass inside the bulge is well determined in the m2m models and it is not sensitive to the dark matter profile. We test the first case using a model with $\Omega_b = 38\freq$, $c_s = 18\kms$. In the lower panel of Figure~\ref{fig:Vrot_comparison} we show the gas velocities of this model projected with $i=77^{\circ}$ and varying inclinations (equation~\ref{eq:iprofile}) using the solid and long dashed curves, respectively. Compared to the upper panel, the velocity peaks at $| V_{los} | \sim 320\kms$ move closer to the center because $\Omega_b$ increases to $ 38\freq$ and produces a smaller nuclear ring. With lower inclinations in the central region ($R < 4\kpc$), the gas velocities are projected to lower values, resulting in $V_{los}$ being more similar to the $\oiii$ data. In our tests, if the inclination profile of the gas model in Figure~\ref{fig:Ob38c18_i_scheme} is allowed to increase sharply by $\sim20^{\circ}$ as the radius shifts inward from $R=0.7\kpc$ to the center, the gas velocities of the model could fit the $\oiii$ data better.

\subsubsection{Twisted zero-velocity lines}
The $\oiii$ velocity field shows a twisted zero-velocity line that crosses most of the shock features. The upper panel of Figure~\ref{fig:ZVL_comparison} presents the region with $| V_{los} | < 2.5\kms$ in our best-fitting model with $i=77^{\circ}$ (black area) and its comparison with the zero-velocity curve in the $\oiii$ data (green curve). The shape of the green curve and the black area are quite similar on the far (lower) side of M31. On the other side, the green curve is found at larger distances from the disk major axis compared to the black area ($0\arcmin < X < 5\arcmin$, or $0 \kpc < X < 1.14 \kpc$), consistent with the asymmetries in the $\oiii$ shock positions. Near the center, the green curve is almost perpendicular to the disk major axis of M31. Such a shape of the zero-velocity line indicates that the gas motion is roughly tangential to the line-of-sight near the minor axis of the disk. On the contrary, the black area shows a twisted shape near the central $3 \arcmin$ ($\sim 680\pc$), indicating that the modeled gas motions are not purely circular. Note that the nuclear ring in our model has a radius of $\sim 1 \kpc$, which is relatively large. In the outer regions of the nuclear ring, the inward gas flow exhibits a deviation from the circular motion, resulting in a slightly elliptical shape. This phenomenon can be seen in the face-on view of our gas model in Figure~\ref{fig:KR241_bestfit_SD_pv}. The deviation from circular motion leads to zero-velocity lines that are slightly shifted from the minor disk axis. This shift can be reduced as the inclination decreases. The difference in the zero-velocity lines between the model and the $\oiii$ data is probably due to the different gas structures near the center, as discussed in \S~\ref{sec:model_varying_inclination}. The zero-velocity line in the model extends to farther regions. The observation coverage of $\oiii$ is limited and does not extend to the two turning points of the black area at $(X,Y) = (-8,-4) \arcmin$ and $(7.3,3.9) \arcmin$. 
Future observations of ionized gas in these regions will help further constrain the models. In the regions outside the bar at $| Y |\sim10 \arcmin$, gas mainly follows circular motions, resulting in zero-velocity lines that are almost parallel to the disk minor axis. Note that our models do not include spiral arms, which could produce weak non-circular motions in the disk. The lower panel of Figure~\ref{fig:ZVL_comparison} presents the same context as the upper panel but for the model with $\Omega_b=38\freq$, $c_s=18\kms$ and varying inclinations. Compared to the upper panel, the central part of the black area is closer to the green curve because the lower inclination there stretch the curve along the $y$-axis. The turning points of the black area shift inward to $(X,Y) = (-5.5,-3.5) \arcmin$ and $(5.5,3.5) \arcmin$ because the shocks in models with higher pattern speeds become shorter. Note that this model with higher pattern speed produces weak spiral features near the outer Lindblad resonance (OLR) ($R_{OLR} \sim 11.2 \kpc$), the non-circular motions of which result in zero-velocity lines slightly deviated from the disk minor axis in the outer region ($| Y |\sim10\arcmin$). 


\subsection{Pattern speed discrepancy between gas and stellar dynamics}
\label{sec:omegab_discrepancy}

By analyzing stellar photometry and kinematics in both the classical bulge region and the boxy/peanut bulge region, \citet{bla_etal_18} concluded that a pattern speed $\Omega_{b}$ of $\sim 40\freq$ is necessary to replicate all observed data. However, from the perspective of gas evidence, the observed shock positions are found far from the bar major axis, which favors a bar with a low pattern speed, even though such a model overpredicts the gas velocities along the disk major axis.

This is not the first time that a discrepancy between gas and stellar dynamics has come up. For the Milky Way, the N-body model of \citet{shen_etal_10} and \citet{Shen_14} determined the bar pattern speed $\Omega_b$ to be $\sim 39\freq$ based on the line-of-sight velocity data, even though the uncertainty range is harder to know. \citet{por_etal_17} first constructed m2m dynamical equilibrium model including a full bulge and a planar long bar. By fitting the Milky Way surveys, they determined the pattern speed to be $\Omega_b = 39 \pm 3.5 \freq$, positioning $\mathcal{R}$ at $1.22\pm0.11$. Kinematic models based on the bar velocity field gave $\sim 41\pm 3\freq$ \citep{san_etal_19}. Later test particle models with a decreasing pattern speed gave $\Omega_b \sim 35\freq$ \citep{chi_etal_21}. More recently, the proper motion comparison of the stellar dynamical m2m model in \citet{por_etal_17} to VIRAC and Gaia data gave $\Omega_b \sim 33 \pm 2\freq$ \citep{cla_etal_22}.  The pattern speed was determined to be $\Omega_b \sim 40 \freq$ through the application of the continuity equation to the kinematics of the updated APOGEE data \citep{leu_etal_23}. For the early gas-dynamical Milky Way models high pattern speeds of $\sim60 \freq$ were found based on comparing preferred features with the (l-v) diagram \citep{eng_ger_99, fux_99, bis_etal_03}. They came out too high because the potential was not well enough known, leading to ambiguity in comparing to the (l-v) diagram. Other models based on different gas observations \citep{wei_sel_99, rod_com_08, sor_mag_15, li_etal_22} gave more correct lower values. This history
suggests that the gas pattern speed is more sensitive to
variations in potential or data. NGC 1365 and NGC 4321 have previously been constrained by gas simulations to be rotating slowly with $\Omega_b\sim20\freq$ \citep{lin_etal_96, zan_etal_08, gar_etal_94}. Applying the Tremaine-Weinberg method to stellar data has given new constraints on the pattern speeds for NGC 1365 and NGC 4321, with $\Omega_b=38.1^{+20.1}_{-21.2}\freq$ and $43.4^{+3.1}_{-9.1}\freq$, respectively \citep{wil_etal_21}.

\subsection{Systematic uncertainty in gas and stellar dynamical models}
\label{sys_uncertainty}

Considering the discrepancy between pattern speeds from stellar dynamics and gas dynamics discussed in \S~\ref{sec:omegab_discrepancy}, it is possible that there are systematic effects due to various assumptions that could come in differently in the two methods.

We cannot rule out the possibility that gas is still perturbed by a recent head-on merger, which could leave the inner gas disk tilted compared to the outer disk. We have investigated if a gas model with varying inclinations could reproduce the shock features and central gas kinematics in \S~\ref{sec:model_varying_inclination} and \S~\ref{sec:other_gas_feature}. Using varying inclinations indeed helps reduce the discrepancy of pattern speeds between gas and stellar dynamical models (see \S~\ref{sec:omegab_discrepancy}). If the inclination of the central gas disk ($R<4\kpc$) is allowed to follow a trend similar to those in $\HI$ and ionized gas observations, the gas model with $\Omega_b=38\freq$ could reproduce most of the gas features. However, it should be noted that we do not intend to find a perfect match with the observed data, but to use this model with varying inclination to verify the possible existence of a tilted inner gas disk.

Although other uncertainties could arise in the gas models, our simulations always favor a low pattern speed $\Omega_b < 30\freq$ if the inclination of gas disk is fixed. We have investigated assumptions like the inclination and the bar angle. As shown in \S~\ref{sec:vary_omegab_cs_I}, although reducing the inclinations of the gas disk helps shift the shocks to the observed positions for models with $\Omega_{b}\sim40\freq$, its SMHD is larger than models with smaller $\Omega_{b}$ by $\sim60\%$ (see the upper panel in Figure~\ref{fig:SMHD_curve_omegab_cs_changeI}). The velocity jumps of the less inclined model with $\Omega_{b}\sim40\freq$ are smaller than the observation in all five slits, especially at $X = -6.0\arcmin$ (see panels a and e in Figure~\ref{fig:SMHD_comparison_7models}). A more end-on bar could help some, but not enough to improve the models with high pattern speeds (see \S~\ref{sec:different_phib} for more details). In addition, the outer in-plane bar is not very clear in the stellar data, so the quadrupole of the bar could be uncertain. We have tested models with different bar quadrupoles in Appendix~\ref{appendix:varying_A2}. Our models prefer a long bar or a large quadrupole scale length for a pattern speed of $\Omega_b = 38\freq$. However, stretching the bar quadrupole outwards does not help improve the high pattern speed model much. The isothermal assumption of EoS may be simple, so we test the more sophisticated simulations that include sub-grid physics in \S~\ref{sec:smuggle_model}. Our results show that even if we consider the multi-phase properties of gas, it still requires a low bar pattern speed of $\sim 18\freq$ to reproduce the observed shock positions. 


For the stellar dynamical model, there is some tension between the photometric and kinematic data, with the latter favoring $\Omega_b=40\pm 5\freq$ and the former preferring lower values. However, the lower $\Omega_b$ inferred from the photometry still encompasses the parameter space region better constrained by the kinematic data. The kinematic data used to constrain the stellar dynamical model is in $V$-band, which is easier to be affected by the dust. On contrast, the IRAC $3.6\micro m$ photometry is less affected by the dust. The difference of $\Omega_b$ derived from photometry and kinematics could be due to the stronger effects of dust on the stellar kinematics. Although the m2m model in \citet{bla_etal_18} could reproduce well the observed asymmetries in the stellar kinematics when including a reasonable dust model, they did not test whether the derived $\Omega_b$ is similar using the symmetrized data between the near and far side of M31. Another possible uncertainty was from the parametrization of the dark matter profile. \citet{bla_etal_18} fitted Einasto profiles that reproduced the bar/bulge kinematics better than the NFW profile, which instead fit $\HI$ circular rotation curves better especially at $R>8\kpc$. The Einasto profile generated the "cored" profile that the data demanded, but it has less space to lower the circular rotation curve and follow the NFW profile in the outer region. Although a possible "cored"-NFW profile could be more flexible, the dynamical masses in the bulge region is well determined. Both the Einasto and NFW profiles result in similar dynamical masses within the bulge region. Note that shock features in our gas models are not sensitive to the dark matter profile as shown in \S~\ref{sec:SMHD_overall_comparison}.

In addition, previous gas simulations have shown that the length and position of shocks depends on many parameters. Lower pattern speeds tend to extend shocks outwards \citep{li_etal_15}; higher sound speeds move shocks inwards and closer to the bar major axis \citep{eng_ger_97, kim_etal_12a}; a larger axis ratio of the bar produces more elongated shocks \citep{kim_etal_12b}. There could have some uncertainty in the degree of boxiness of the boxy/peanut bulge in the m2m models. Better constrained parameters in the m2m models like the bar height to length ratio and the dark matter profile outside the bar may help us improve the gas models with large pattern speeds of $\sim 40\freq$. 

Previous magnetohydrodynamical simulations of barred galaxies have shown that as the magnetic field becomes stronger, the size of the nuclear ring decreases and the shock feature move inwards \citep{kim_sto_12}. If we include magnetic field in our models, this would require an even lower bar pattern speed to fit the observed shock features. On the other hand, \citet{moo_etal_23} investigated the effects of magnetic field on the nuclear ring in barred galaxies and found that the existence of magnetic field largely suppresses the star formation on the nuclear ring and helps produce a circumnuclear disk. This may help to explain the relatively weak star formation activity in the central $500\pc$ of the M31 bulge \citep{dong_etal_18}. A detailed study about the effects of magnetic field on gas features in M31 is beyond the scope of this paper. Nevertheless, precise measurements on the magnetic field in M31 would help construct more sophisticated gas models in the future.

\subsection{Slow and large bars in other galaxies}

Large bars with slow pattern speed are not anomalies in observations. \citet{gar_etal_22} used the Tremaine-Weinberg (TW) method to measure the bar pattern speed for a sample of 97 MW-analogue galaxies in MANGA. They found that 52 galaxies host slow bars with $\mathcal{R} > 1.4$. The also showed that longer and more massive bars tend to rotate more slowly, which is consistent with \citet{cuo_etal_20}. A similar study by \citep{ger_etal_23} suggested 62 percent of the barred galaxies in the galaxy zoo project are identified to have slow bars with $\mathcal{R} > 1.4$ based on TW analysis. They also found that stronger bars tend to have lower pattern speeds. More recently, Zou et al. (in preparation, private communication) used a larger sample of 174 nearby galaxies in MANGA and found that bar pattern speed decreases when the bar length increases, but the authors note this relation may simply result from a nearly constant $\mathcal{R}$ ratio.


A bar with a pattern speed of $\sim 18\freq$ has been found in MW-analogue galaxies. All the 97 MW-analogue galaxies in \citet{gar_etal_22} have stellar masses $M_{\star}$ within the range of $10^{10.3}-10^{11.3} \Msun$. Note that M31 is determined to have a total stellar mass of $10^{11}-10^{11.18} \Msun$ \citep{tam_etal_12}, which lies within the upper limit of the sample stellar mass. The bar pattern speeds of the sample in \citet{gar_etal_22} have a smooth distribution of $\Omega_b = 28.14^{+12.30}_{-9.55}\freq$, with $\mathcal{R} = 1.35^{+0.60}_{-0.40}$. The pattern speed of our best-fitting model of $\sim 18\freq$ lies within the lower $1\sigma$ limit of their distribution. However, it should be noted that the most massive galaxies in their samples have large bar size $a_{bar}$ over $6 \kpc$. The median of bar pattern speeds in their samples with $a_{bar} \sim 4 \kpc$ is around $35\freq$. With the constraints of the bar size and the disk rotation curve ($a_{bar} < 6\kpc$, $190 < V_{c} < 290 \kms$), they further obtained a sub-sample of 25 MW-analogue galaxies with $\Omega_b = 30.48^{+10.94}_{-6.57}\freq$ and $\mathcal{R} = 1.45^{+0.57}_{-0.43}$. The pattern speed of the Milky Way $\Omega_b \sim 40 \freq$ lies close to the upper $1\sigma$ limit of this distribution. Most slow bars in their 97 MW-analogue galaxies have $\mathcal{R}< 1.7$. It is also uncommon to see bars with $\RCR=14.1 \kpc$, $\mathcal{R}> 3.0$ in galaxies with stellar masses similar to M31 in large surveys \citep{fon_etal_17, guo_etal_19, gar_etal_20, gar_etal_22, lee_etal_22}. Using a sample of 174 galaxies with total stellar masses in the range of $10^{9}-10^{10.92}\Msun$, Zou et al. (in preparation) showed that there is no clear trend between pattern speeds and galaxy stellar masses. The pattern speed of their sample falls within the interval of $10$ to $45\freq$, with a median value approximately around $30\freq$.


Barred galaxies with $\mathcal{R} > 3$ are not common, but the result we got using $i=77^{\circ}$ is not unreasonable. $\mathcal{R}$ is very sensitive to the bar length, which is not easy to determine and could have some uncertainty. If the bar length in M31 model is allowed to extend to $\sim 5 \kpc$ and rotate a bit faster at $\Omega_b = 22\freq$, then the $\mathcal{R}$ ratio will be positioned at $2.32$. The bar parameters would not be that extreme.

\subsection{Further improvement in a tidal interaction scenario}
\label{sec:tidal_interaction_scenario}
In this work, our main focus is the shock features inside the bar region of M31. However, there is another intriguing structure in M31, the 10-kpc ring, which is rich in gas and dust. According to \citet{lew_etal_15}, this ring may have been created by the OLR resonance of a central bar and later perturbed by a recent head-on collision. Our current model does not have a 10-kpc ring so if this is correct an alternative mechanism is required, such as the suggested merger event. The m2m models in \citet{bla_etal_18} suggest that a bar pattern speed of $\Omega_{b} \sim 40\freq$ is required to position the OLR at $R\sim 10\kpc$. However, our analysis reveals that if the bar rotates at a pattern speed of $\Omega_{b} \sim 40\freq$, the bar-driven shock features will not match the $\oiii$ data, unless the inner gas disk has a smaller inclination than the stellar disk. On the other hand, if the 10-kpc ring in M31 is indeed the result of a recent merger event as proposed by \citet{blo_etal_06}, it remains unclear how much the collision has affected the central gas kinematics. Moreover, it is possible that the central bar has produced the 10-kpc ring, and a head-on collision has subsequently altered the inclinations of the central gas features and perturbed the outer ring. To gain a better understanding of these questions and the formation history of M31, further investigations using an interaction model are necessary.

\section{Conclusion}
\label{sec:conclusion}

We have run a series of high-resolution gas simulations that use realistic gravitational potentials of M31. These potentials are derived from the made-to-measure (m2m) models in \citet{bla_etal_18}, which are well-constrained using stellar photometry and kinematics in the bulge and disk regions of M31. Our gas simulations allow us to independently constrain the pattern speed of the bar in M31 by fitting the observed shock features in the bulge region. Our findings are summarized as below:

(1) Our best-fitting models, with a low pattern speed of $\Omega_{b} = 18\freq$ and a fixed inclination of $i=77^{\circ}$, reproduce the observed shock features in $\oiii$ data reasonably well (\S~\ref{sec:define_goodness_of_fit}, Figure~\ref{fig:KR241_bestfit_SD_pv}). Larger effective sound speeds over 30 $\kms$ even improve the fittings. However, the best-fitting models overpredict the velocities on the disk major axis, and the zero-velocity lines in the central $2\arcmin \times 2\arcmin$ ($\sim 460\pc \times 460 \pc$) are not aligned with the one in $\oiii$ data. 

(2) We simulated gas models with various m2m potentials. Although these models differ slightly in their goodness of fit, they all support a relatively low pattern speed in the range of $16-20\freq$ (\S~\ref{sec:SMHD_overall_comparison}, Figure~\ref{fig:overview_SMHD_I77}).

(3) Observations of ionized and neutral gas have indicated that the inclination of the inner gas disk of M31 is lower than that of the outer disk. However, the exact 3D structure of the inner gas disk is still uncertain. Decreasing the inclination of the gas disk helps improve the goodness of fit for models with higher pattern speeds. Notably, for inclinations at $i \geq 69^{\circ}$, the SMHD results favor a pattern speed of $\Omega_{b} \leq 30 \freq$ (\S~\ref{sec:SMHD_overall_comparison}, Figure~\ref{fig:overview_SMHD_changeI}). We also examined various models for the variation of inclination inside $\sim20 \arcmin$ ($4.6\kpc$) and found that the shock features favor higher pattern speeds as the central inclination decreases (\S~\ref{sec:model_varying_inclination}).
    
(4) Since the interstellar medium in the real universe is multi-phase, we explored a more sophisticated model with SMUGGLE module turned on in the standard AREPO code for constant inclination $i=77^{\circ}$. Although the shock features of this model are smoother than those in the isothermal models, they also require a low pattern speed of $\Omega_{b} \sim 18\freq$ to produce shocks at positions similar to the $\oiii$ data (\S~\ref{sec:smuggle_model}, Figure~\ref{fig:Ob18_smuggle_fit_SD_pv}).

(5) If the inclination of the central gas disk is allowed to follow a trend similar to the observations of $\HI$ and ionized gas, the gas model with $\Omega_b = 38\freq$ can both match the shock features (\S~\ref{sec:model_varying_inclination}, Figure~\ref{fig:Ob38c18_varyingi_SD_pv}), the velocities on the disk major axis, and the zero-velocity line (\S~\ref{sec:other_gas_feature}, Figures~\ref{fig:Vrot_comparison} and~\ref{fig:ZVL_comparison}). In this scenario the pattern speeds in gas and stellar dynamical models are more consistent.

Despite being our nearest galaxy, M31 still has many mysteries that remain to be explored. This study has found supporting evidence for an inner tilted ring within the central $1\kpc$ of the gas disk, however the exact cause of it remains uncertain. \citet{ali_etal_23} predicted the evolution of inner gas streams within central $500\pc$ of the M31 potential. In their simulations a hot gas atmosphere with temperatures around $10^6\;K$ is required for the formation of a nuclear spiral. This central region has not been well resolved in the present simulations. Beyond this, a better understanding for the mechanisms behind the 10-kpc ring is needed, even if previous studies have proposed various theories including the influence of the bar's OLR \citep{bla_etal_18}, minor mergers \citep{blo_etal_06}, or gas accretion during major mergers \citep{ham_etal_18}. Exploring these mechanisms for the origin of the outer 10-kpc ring as well as the inner tilted ring would require tailored N-body and gas-dynamical simulations of galaxy interaction models. In this study, we primarily used the barred gravitational potential from stellar dynamical models to construct gas models, which enabled us to gain insights into various gas features in M31. Nevertheless, the question of whether the bar in M31 formed through secular evolution or galaxy interactions remains a subject for future investigation.


\software{
{\tt Athena++} \citep{sto_etal_20},
{\tt AREPO} \citep{volker_10, wei_etal_20},
NumPy \citep{2020NumPy-Array},
SciPy \citep{2020SciPy-NMeth},
Matplotlib \citep{4160265},
Jupyter Notebook \citep{Kluyver2016jupyter}
}

\begin{acknowledgments}

The authors would like to thank Laurent Chemin for sharing $\HI$ data of M31. The research presented here is partially supported by the National Key R\&D Program of China under grant No. 2018YFA0404501; by the National Natural Science Foundation of China under grant Nos. 12103032, 12025302, 11773052, 11761131016; by the ``111'' Project of the Ministry of Education of China under grant No. B20019; and by the China Manned Space Project under grant No. CMS-CSST-2021-B03.  J.S. also acknowledges support from a \textit{Newton Advanced Fellowship} awarded by the Royal Society and the Newton Fund. M.B. acknowledges funding from ANID through the FONDECYT Postdoctorado 2021 Nr 3210592 and the Excellence Cluster ORIGINS founded by the Deutsche Forschungsgemeinschaft (DFG; German Research Foundation) under Germany’s Excellence Strategy – EXC-2094 – 390783311. This work made use of the Gravity Supercomputer at the Department of Astronomy, Shanghai Jiao Tong University, and the facilities of the Center for High Performance Computing at Shanghai Astronomical Observatory. 
\end{acknowledgments}

\clearpage 


\bibliographystyle{aasjournal}
\bibliography{ms_submit}
\appendix
\twocolumngrid

\setcounter{figure}{0}
\renewcommand{\thefigure}{A\arabic{figure}}

\section{Comparison between 2D isothermal and more sophisticated models}
\label{appendix:eos_comparison}

\begin{figure*}[t!]
\includegraphics[width=\textwidth]{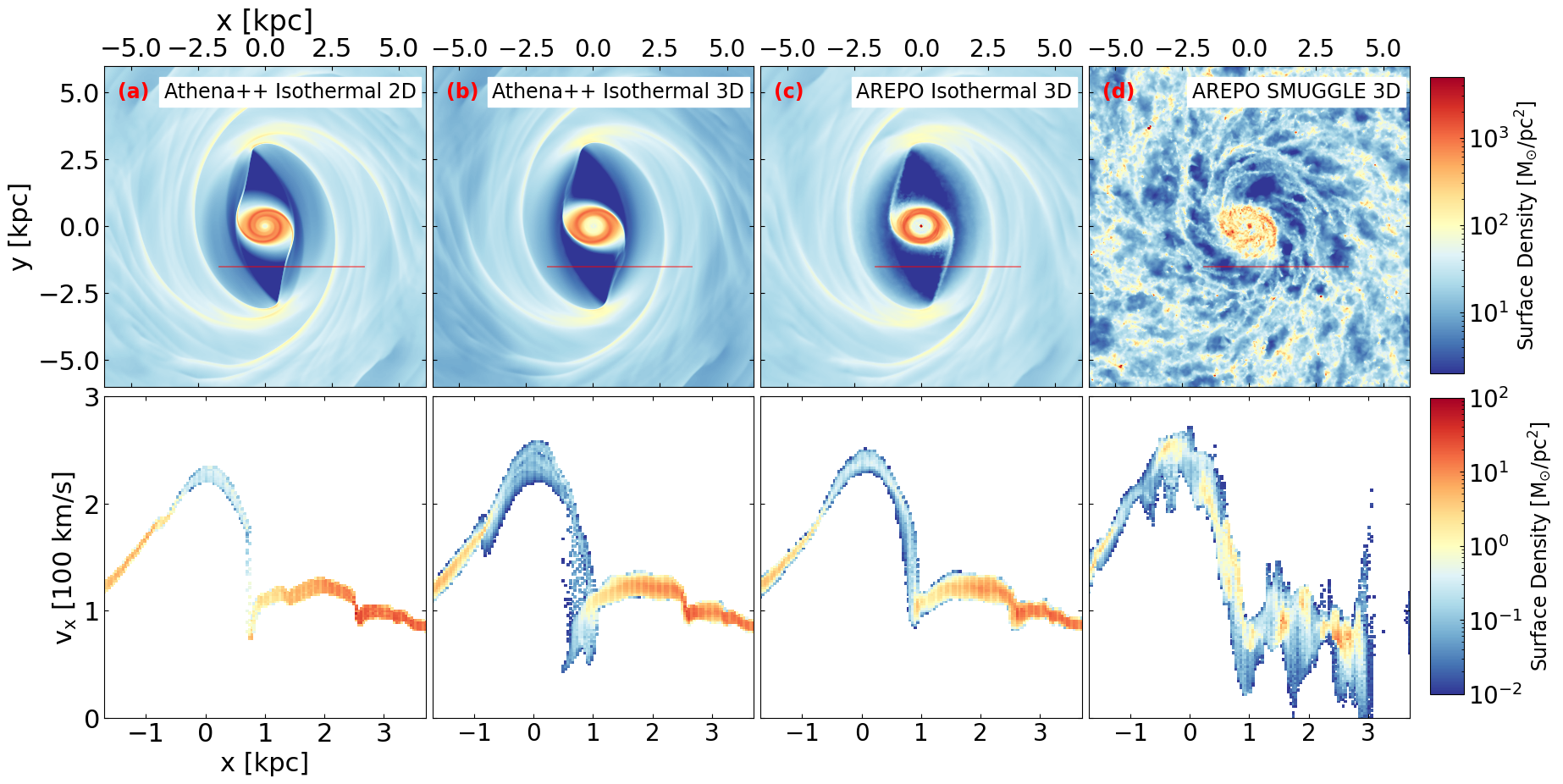}
\centering
\caption{Gas surface density (upper panels) and kinematics (lower panels) at T = 525 $\Myr$ of a test simulation using (a) Athena++ 2D model, (b) Athena++ 3D model, (c) AREPO 3D model, and (d) AREPO-SMUGGLE 3D model. All models use a bar pattern speed of $\Omega_{b} = 38 \freq$. Models in (a), (b), and (c) use the isothermal equation of state with an effective sound speed of 15 $\kms$. The model in panel (d) turns on heating and cooling processes, self-gravity, star formation, and stellar feedback. Although this model shows a more filamentary morphology, the main gas streams can still be recognized. The red line segments in the upper panels indicate pseudo-slits crossing the shocks almost perpendicularly. The velocity distribution along the pseudo-slit is measured in the lower panels. The overall profile of gas kinematics of these four models is similar.}
\label{fig:gas_SD_pv_different_code}
\end{figure*}

To compare the gas pattern and kinematics between the 2D isothermal model and the more sophisticated models, we construct four models as follows: 
    
$-$ Model 1: a 2D isothermal model with Athena++ code; 
    
$-$ Model 2: a 3D isothermal model with Athena++ code; 
    
$-$ Model 3: a 3D isothermal model with the standard AREPO code; 
    
$-$ Model 4: a 3D multiphase model with SMUGGLE module turned on in the AREPO code. 
    
Model 2 is set up with a $2048 \times 2048 \times 53$ grid, covering a length of $0.78\kpc$ along the vertical direction. The initial settings of the models are similar to those introduced in \S\ref{sec:initial_setting}, except that we use a different scale height of $z_{\rm gas} = 0.1 \kpc$ for the initial gas disk. Models 1, 2, and 3 use a sound speed of $c_{s} = 15\kms$, which has been commonly used in previous studies \citep{kim_etal_12a}. All models use a bar pattern speed of $\Omega_b = 38\freq$ and a potential of the JR804 m2m model. 

The upper panels in Figure~\ref{fig:gas_SD_pv_different_code} present the gas surface density of the above models. Although there are small differences, the overall gas pattern in the three isothermal models (panels a, b, and c) are similar. Model 4 (panel d) shows more complicated gas streams, including the filaments and holes produced by self-gravity and stellar feedback. Despite the increased complexity, the main gas substructures, such as the nuclear ring, off-axis shocks, and bar-driven spirals, can still be identified in positions similar to those in the isothermal runs. The lower panels in Figure~\ref{fig:gas_SD_pv_different_code} present the gas velocities $v_{x}$ along pseudo-slits perpendicular to the shocks (red line segments in the upper panels). The overall gas PVDs of the four models are similar, showing large velocity jumps at the shock region. The velocity jump feature at $X \sim 3\kpc$ in panel (d) is produced by a transient supernova explosion and will disappear in a few $\Myr$. While star formation and stellar feedback regulate local gas properties, the large-scale gravitational potential primarily determines the overall gas features on the PVDs. Therefore, we can consider the Athena++ 2D isothermal models as a first-order approximation of the gas observations in M31.

\section{FAST observation of the HI disk in M31}
\label{appendix:HI_obs}
The data were observed in drift scanning with the 19-beam L-band receiver, which has a beam size of approximately $2.9\arcmin$ at 1.4 $\rm GHz$ \citep{jia_etal_20}. To calibrate the data, a noise diode with about 10 K was injected every 5 minutes for a duration of 2 seconds during observation, along with observations of the quasar 3C48 as a flux calibrater. The data was processed using HiFAST \footnote{\url{https://hifast.readthedocs.io}} (Jing et al., in preparation), and the final data cube produced had a pixel size of $1\arcmin$ and a velocity resolution of 1.61 $\kms$. To capture the velocity information of the inner gas disk, we position a pseudo-slit along the disk major axis with a width of $40''$ and generate the $\HI$ PVD (see \S~\ref{sec:other_gas_feature}). To reduce contamination from the outer HI warp (lower velocity features on the PVD), we extract the outer envelope of the $\HI$ PVD. 

\section{Effects of varying the strength and length of bar}
\label{appendix:varying_A2}

Except for the pattern speed, the property of the gravitational potential of the bar is one of the most important parameters to determine the properties of shocks \citep{athana_92}. \citet{sor_etal_15c} systematically explored the effects of varying the quadrupole component of the gravitational potential on the gas flows. They extracted the quadrupole component $\Phi_{2}$ using the Fourier decomposition and adjusted its strength $A$ and scale length $r_{q}$. In their simulations, the gas morphology changes a lot as $A$ increases, producing smaller nuclear rings and more elongated shocks. $r_{q}$ mainly affects the strength of spiral arms. A larger $r_q$ strengthens the spiral arms by extending the quadrupole. The increases of $A$ and $r_{q}$ affect the gas kinematics as well, enhancing the non-circular motions in regions of bars and spirals, respectively. 

The observed shock features of $\oiii$ extend almost to the end of the bar (bar half-length equals $L_{bar}=4\kpc$ in m2m models), and are far from the bar major axis. It requires either a low pattern speed to move resonances and gas substructures outwards, or a weaker and longer bar to produce shocks both extended and far from the bar major axis. Based on JR804 potential, we construct several models with adjusted quadrupoles and compare their shock features with the observed ones. We first expand the gravitational potential in multipoles: 

\begin{figure*}[t!]
\centering
\includegraphics[width=\textwidth]{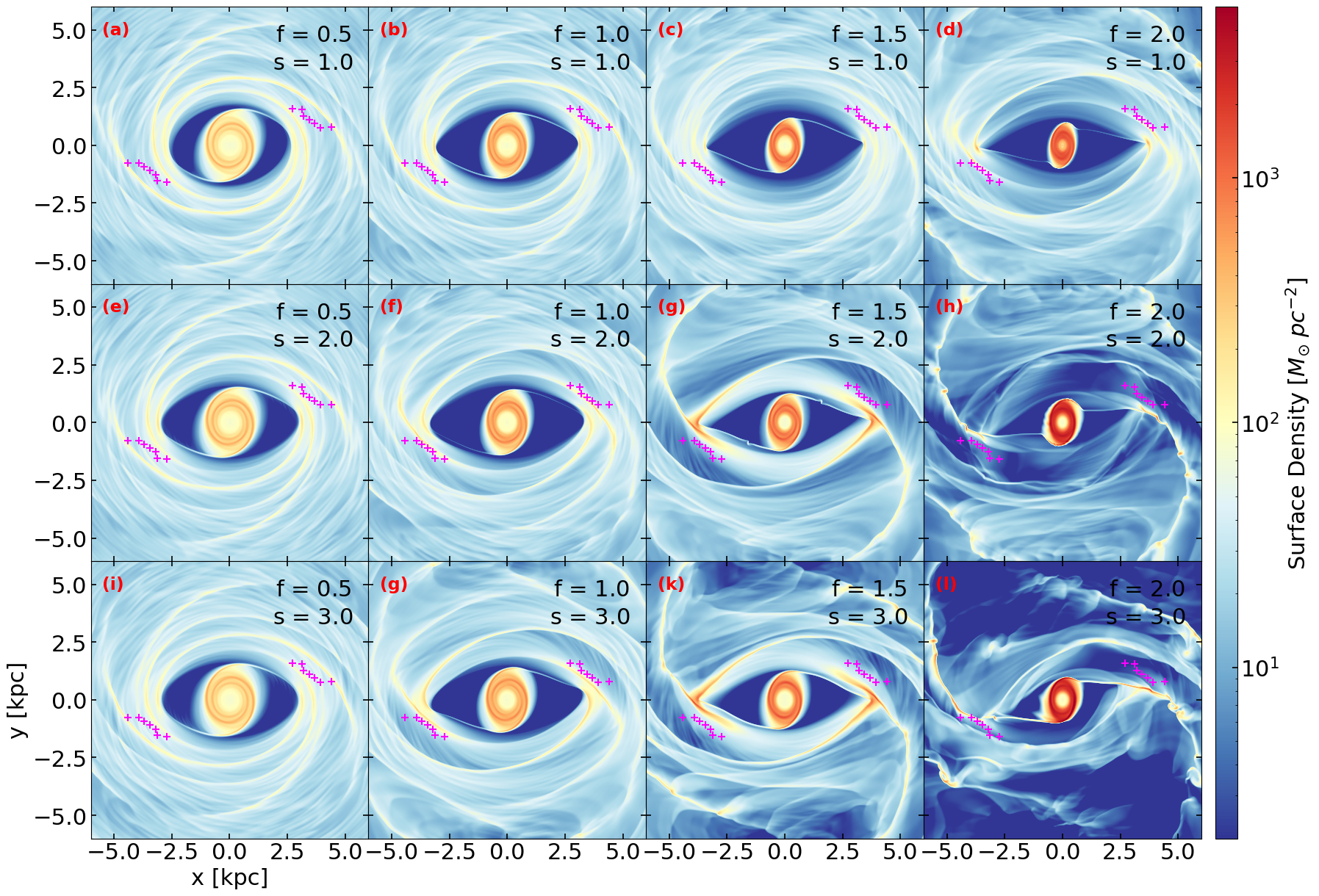}
\caption{Gas surface density in models with different adjusted quadrupoles. $f$ and $s$ are two scaling factors determining the strength and the scale of quadrupoles, respectively. $f$ is increasing from the left to the right with values of 0.5, 1.0, 1.5, and 2.0. $s$ is increasing from the top to the bottom with values of 1.0, 2.0, and 3.0. $f$ and $s$ are defined in equation (B3). The bar major axis is aligned with the $x$-axis. The snapshots are taken at $T = 800\Myr$. All models use a pattern speed of $\Omega_{b} = 38\freq$ and a sound speed of $c_{s} = 10 \kms$. The highlighted plus signs represent the positions of $\oiii$ shock features on the far side of M31 (data points on the near side are the mirrored version of the far side), deprojected to the face-on view with an inclination of $77^{\circ}$ and a bar angle $\phi_b$ of $54.7^{\circ}$.}
\label{fig:12fs_pureA2_SD_comparison_Ob38c10}
\end{figure*}

\begin{equation}
\Phi(R,\phi) = \Phi_0(R)+\sum_{m=1}^{\infty}\Phi_{m}(R)cos(m\phi+\phi_{m}),
\end{equation}

here $\phi_{m}$ are constants. We assume that $\phi_{2} = 0$. The $\Phi_{2}$ can be extracted using the Fourier decomposition: 

\begin{equation}
\Phi_{2}(R) = \dfrac{\int_0^{2\pi}\Phi(R,\phi)cos(2\phi)d\phi}{\pi}.
\end{equation}

We adjust the profiles of $\Phi_{2}(R)$ into $\Phi_{2}(R)'$ by : 

\begin{equation}
\Phi_{2}(R)' = f\Phi_{2}(R/s),
\end{equation}

here $f$ and $s$ are two scaling factors, determining the strength and the scale of $\Phi_{2}(R)'$, respectively. 

We start with the fiducial pattern speed of $\Omega_{b} = 38 \freq$ of the JR804 model and an effective sound speed of $10\kms$. Figure~\ref{fig:12fs_pureA2_SD_comparison_Ob38c10} presents the gas patterns with different $f$ and $s$. The highlighted plus signs represent the shock positions in $\oiii$ deprojected with an inclination of $77^{\circ}$ and a bar angle of $54.7^{\circ}$. For those models that are bisymmetric but the observed shock features are asymmetric, we only select the shock features on the far side as a reference. The shocks become more elongated as $f$ increases from 0.5 to 2, similar to the results in \citet{sor_etal_15c}. A weaker quadrupole produces a larger nuclear ring and shocks more away from the bar major axis, even though the shock is too short compared to the deprojected shock positions. For models with stronger quadrupoles in the third and fourth columns, although the length of shocks is similar to those in observations, the shocks are too close to the bar major axis. The effects of $s$ are hard to see for the fiducial strength of quadrupoles ($f = 1$), but they are clear to see for the enhanced quadrupoles ($f = 2$). The quadrupole length almost extends to the corotation radius of the bar of $R_{CR}\sim6.4\kpc$ with $\Omega_{b} = 38 \freq$ in the bottom row. In panels (h) and (l), the bar-driven spirals are significantly strengthened and become shocks. However, the shock positions are roughly similar to the one in panel (d). Note that the bar-driven spirals are different from the shocks inside the bar, and we do not intend to use bar-driven spirals to match the observed shock features. According to the results of these tests, we propose that a slightly different quadrupole component of the bar does not help improve the fitting with the fiducial pattern speed of $\Omega_b = 38\freq$. 

\section{Models with different masses of classical bulge}
\label{appendix:varying_MCB}

\begin{figure}[t!]
\includegraphics[width=\columnwidth]{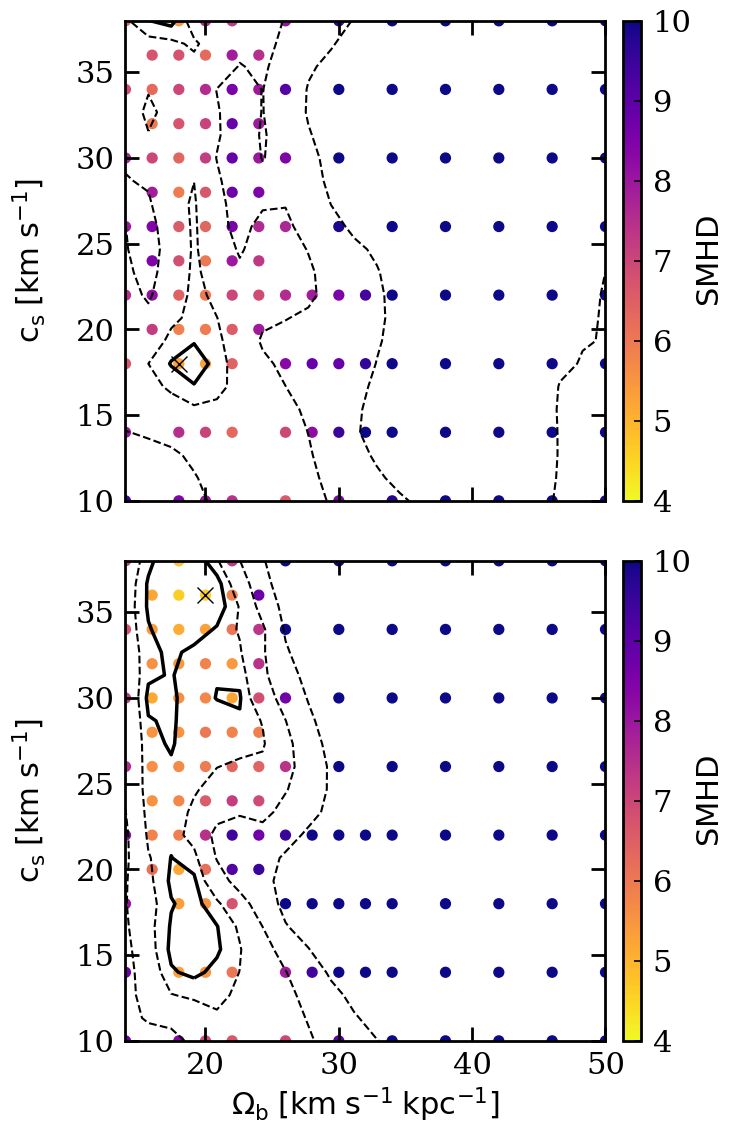}
\caption{Same context as the left panels of Figure~\ref{fig:overview_SMHD_I77} but for models with JR644 (top) and JR724 (bottom) potentials.}
\label{fig:overview_SMHD_I77_644_724}
\end{figure}

Previous gas simulations revealed that the central density of the bulge affects significantly the shape and size of the nuclear ring \citep{li_etal_15}. In our control models, the gas models without a classical bulge component produce the $x_{1}$ type nuclear rings, resulting in shocks very close to the bar major axis. A massive classical bulge component is necessary to provide the centrifugal force for the $x_{2}$ type nuclear rings. As the masses of the classical bulge increase, the nuclear ring sizes increase slightly. Nevertheless, the masses of the classical bulge mainly affect the central gas morphology and do not alter the positions and kinematics of shocks. 

The m2m models with smaller pattern speeds host less massive classical bulges and more massive boxy/peanut bulges, as listed in table~\ref{tab:par_m2m_models}. To see if a different mass of classical bulge changes the fitting results, we construct a series of gas models with JR644, JR724, and JR804 potentials, using $M_{\star}^{CB} = 1.02$, 1.1, and $1.18\times10^{10} \Msun$ respectively. The masses of dark inside the bulge are the same in all of these models with $M_{DM}^{B} = 1.2\times10^{10} \Msun$. The upper and lower panels of Figure~\ref{fig:overview_SMHD_I77_644_724} present the results of SMHD in gas models with JR644 and JR724 potentials, respectively. Although the overall distributions of SMHD in Figure~\ref{fig:overview_SMHD_I77_644_724} and ~\ref{fig:overview_SMHD_I77} are slightly different, the fittings of these gas models all favor a low pattern speed of $\Omega_{b} \sim 20 \freq$. Therefore we propose that a different mass of the classical bulge within the range of $93-107\%$ of the fiducial value of $M_{\star}^{CB} = 1.1\times10^{10} \Msun$ does not change our main result. 

\end{document}